\documentclass[aip,reprint,jcp]{revtex4-1} 
\usepackage[%
  colorlinks=true,
  urlcolor=blue,
  linkcolor=blue,
  citecolor=black
]{hyperref}
\usepackage{graphicx}
\usepackage{xcolor}
\usepackage{dcolumn}
\usepackage{bm}
\usepackage[utf8]{inputenc}
\usepackage[T1]{fontenc}
\usepackage{mathptmx}
\usepackage{etoolbox}
\usepackage{amsmath}
\renewcommand{\selectlanguage}[1]{}
\DeclareUnicodeCharacter{2009}{\,} 
\DeclareUnicodeCharacter{2032}{'}

\makeatletter
\def\@email#1#2{%
 \endgroup
 \patchcmd{\titleblock@produce}
  {\frontmatter@RRAPformat}
  {\frontmatter@RRAPformat{\produce@RRAP{*#1\href{mailto:#2}{#2}}}\frontmatter@RRAPformat}
  {}{}
}%
\makeatother

\begin{document}
\preprint{AIP/123-QED}

\def\*#1{\mathbf{#1}}
\def\&#1{#1}

\title{Mechanisms of Chain Exchange in Diblock Copolymer Micelles}

\author{Samuel Varner}
\affiliation{Division of Chemistry and Chemical Engineering, California Institute of Technology, Crellin Laboratory, Pasadena, CA 91125, United States}

\author{Marcus M\"{u}ller}
\affiliation{Institut für Theoretische Physik, Georg-August-Universität Göttingen, Friedrich-Hund-Platz 1, D 37077 Göttingen, Germany}

\author{Alejandro Gallegos}
\affiliation{Department of Chemical and Materials Engineering, New Mexico State University, Las Cruces, New Mexico 88001, United States}

\author{Timothy P. Lodge}
\affiliation{Department of Chemical Engineering \& Materials Science, University of Minnesota, Minneapolis, Minnesota 55455, United States}
\affiliation{Department of Chemistry, University of Minnesota, Minneapolis, Minnesota 55455, United States}

\author{Kevin D. Dorfman}
\affiliation{Department of Chemical Engineering \& Materials Science, University of Minnesota, Minneapolis, Minnesota 55455, United States}

\author{Zhen-Gang Wang}
\affiliation{Division of Chemistry and Chemical Engineering, California Institute of Technology, Crellin Laboratory, Pasadena, CA 91125, United States}
\email{zgw@caltech.edu}

\date{\today}

\begin{abstract}
We investigate the mechanism of chain exchange in diblock copolymer micelles using two distinct yet complementary simulation techniques. First, the spectral adaptive biasing force enhanced sampling method is combined with coarse-grained molecular dynamics to compute a two-dimensional free energy surface for the chain expulsion process in the strong segregation regime. To facilitate chain expulsion, a distance-based collective variable is biased, and the end-to-end distance of the core block is additionally biased to ensure sufficient sampling of chain conformations. The resulting free energy surface reveals a bimodal distribution of chain conformations along the effective reaction coordinate. The minimum free energy pathway, computed via the string method, qualitatively aligns with the Halperin--Alexander budding-like mechanism. The free energy barrier along this pathway is calculated for core block lengths ranging from $N_\text{core}=4$--$100$, and the barrier is shown to scale as $\beta\Delta F_\text{barr} \sim N_\text{core}^{2/3}$, consistent with the Halperin--Alexander prediction for a globular transition state. Notably, the free energy surface also reveals a nearly degenerate alternative pathway in which the chain escapes by extending out “bead-by-bead,” in agreement with previous simulations. We also study the case of a dense copolymer melt, where the core-block shrinks but does not collapse into a dry compact globule in the opposite phase. To examine the kinetic pathway, a simplified model is introduced in which a single chain escapes from a planar interface within a mean-field background. Using Monte Carlo moves to drive forward flux sampling simulations, the unbiased exchange rate and corresponding free energy barrier are computed. These calculations yield a linear scaling of the barrier, $\beta\Delta F_\text{barr} \sim N_\text{core}$, in agreement with experimental observations and prior simulations. Moreover, analysis of successful escape trajectories highlights an extended chain conformation at the transition state, providing further evidence that experimental conditions favor the hyperstretching escape mechanism over the Halperin--Alexander mechanism.
\end{abstract}

\maketitle

\section{ \label{sec:level1} Introduction}
Block copolymers (BCPs) are amphiphilic molecules that can self-assemble into nanostructured materials in both melts and solutions. In solution, BCPs can spontaneously self-assemble into micellar structures of various morphologies, such as spheres, cylinders, and vesicles depending on the relative volume fraction of the two blocks and their degrees of incompatibility with the solvent and with each other.\cite{zhulina_diblock_2005} The thermodynamic and kinetic properties of copolymer micelles including their size and stability under environmental changes naturally inform their use in applications such as nanoreactors,\cite{cotanda_functionalized_2012,khullar_block_2013,boontongkong_cavitated_2002,bakshi_colloidal_2014} drug delivery and encapsulation,\cite{kazunori_block_1993,luo_cellular_2002,kim_overcoming_2010,kataoka_block_2012,gaucher_block_2005,chiappetta_polyethylene_2007} and nanolithography. \cite{forster_amphiphilic_1998,lohmuller_nanopatterning_2011}

At concentrations exceeding the critical micelle concentration (CMC), micelles form through a two-stage mechanism, starting with rapid formation of small aggregates, and ending with slow equilibration of the micelle size distribution and micelle concentration.\cite{aniansson_kinetics_1974,aniansson_kinetics_1975,aniansson_theory_1976,zana_dynamics_2005,goldmints_micellar_1997} Aggregate formation is fast in concentrated solutions due to a low aggregation free energy barrier, which has been thoroughly discussed in many studies of BCP micelle kinetics.\cite{dormidontova_micellization_1999,semenov_dynamics_2002} In contrast, equilibration is slow due to the large free energy barriers associated with available mechanisms, including single-chain (or unimer) exchange, and fission/fusion.\cite{lodge_dynamics_2022} In single-chain exchange, a chain escapes from one micelle, diffuses through solution, and enters another. Fusion occurs when two aggregates or micelles (possibly of different sizes) combine to form a larger micelle, while fragmentation is its reverse. Although this work focuses on single-chain exchange, a recent review provides a comprehensive discussion of BCP micelle dynamics and equilibration, including open questions in the field.\cite{lodge_dynamics_2022}


The importance of exchange in micelle equilibration has motivated extensive theoretical and experimental studies of its mechanism and free energy barrier. The earliest and most widely cited theory was derived by Halperin and Alexander in 1989, who proposed a free energy barrier for single-chain exchange.\cite{halperin_polymeric_1989,halperin_micellar_2011} They also considered fusion and fragmentation but concluded that these processes are much slower near equilibrium.\cite{nyrkova_theory_2005} Exchange is rate-limited by the barrier for a chain to escape from the soluble micelle core into the less favorable solvent.\cite{cao_fluorescence_1991} For BCPs with much longer hydrophobic blocks than conventional nonionic surfactants, this barrier can reach hundreds of $k_B T$, effectively prohibiting equilibration on experimental timescales.\cite{creutz_exchange_1997,creutz_dynamics_1998} Halperin and Alexander postulated that the chain escapes from the micelle with the core block in a collapsed state to minimize unfavorable contacts with the solvent and corona. Their mechanism applied with Kramers' rate theory led to an escape time of $\tau_\text{esc}\sim \exp(\gamma \rho^{-2/3} N_\text{core}^{2/3}/k_B T)$, where $\gamma$ is the interfacial tension, $\rho$ is the segment density, and $N_\text{core}$ is the degree of polymerization of the core-forming block. The scaling of $N_\text{core}^{2/3}$ follows directly from their assumption of a compact spherical globule. The prefactor to the exponential includes the dependence on the corona-forming block, scaling as $N_\text{corona}^{9/5}$ for star-like micelles and $N_\text{corona}^{7/3}$ for crew-cut.

In the limit of melts with a large invariant degree of polymerization, $\bar{\mathcal{N}}$, however, a single core block embedded in a matrix does not collapse; its relative size change only scales as $\Delta R/R \sim \chi N / \sqrt{\bar{\mathcal{N}}}$. \cite{Muller1998Dec} For a fully solvated core block, the associated free-energy barrier is given by $\Delta F_\text{melt} / k_B T \sim \chi N f_\text{core}$.
In contrast, the Halperin–Alexander scenario \cite{halperin_polymeric_1989} predicts $\Delta F_{\rm HA} \sim \gamma (N_\text{core}/\rho)^{2/3} \sim k_BT \bar{\mathcal{N}}^{1/6}f_\text{core}^{2/3}\sqrt{\chi N}$, where $\rho$ is the segment density, and we have used the strong-segregation estimate for the interfacial tension $\gamma$. \cite{Helfand1972Apr} Consequently, for long-chain melts, the ratio of the two barrier estimates is
$\Delta F_\text{melt}/\Delta F_\text{HA} \sim \sqrt{\chi N}/\bar{\cal N}^{1/6}<1$, implying that the free-energy barrier scales linearly with $N_\text{core}$.\cite{muller_dynamics_2021}

For decades, fluorescence and non-radiative energy transfer experiments were analyzed according to the Halperin--Alexander theory,\cite{prochazka_nonradiative_1991,cao_fluorescence_1991,wang_exchange_1995,smith_determination_1996,creutz_exchange_1997,creutz_dynamics_1998, underhill_chain_1997, rager_micelle_1999, van_stam_tuning_2000} until the development of time-resolved small-angle neutron scattering (TR-SANS) enabled more direct measurements of the exchange rate, and therefore the exchange free energy barrier.\cite{willner_time-resolved_2001,won_molecular_2003} TR-SANS elucidates the exchange rate by tracking the decay of scattering intensity as chains hop between micelles in a solution containing two distinct micelle populations (normal versus perdeuterated cores). By selecting a solvent with a scattering length density intermediate between the two core types, the exchange rate is directly related to the decay of the scattering intensity, $R(t)$, as the chains mix over time. Measurements at several temperatures can be combined into a master curve using time–temperature superposition, extending the dynamic range of TR-SANS over 12 decades.\cite{lu_remarkable_2015} Exchange has been shown to follow rate-limited kinetics with an activation barrier, which should yield a single-exponential decay of scattering intensity with time.\cite{aniansson_kinetics_1974,aniansson_kinetics_1975,halperin_polymeric_1989} TR-SANS measurements, however, consistently revealed a broad relaxation much closer to a logarithmic decay. The explanation for this anomalous behavior is both simple and elegant: because the exchange rate is essentially the escape rate, which depends exponentially on the core-block length, a broad relaxation arises naturally from the polydispersity of the core block. Lund and co-workers were the first to address the role of polydispersity,\cite{lund_logarithmic_2006,lund_equilibrium_2006,lund_unraveling_2007} and Choi and co-workers subsequently connected it directly to the logarithmic decay.\cite{choi_mechanism_2010} Zinn et al. later confirmed this interpretation by demonstrating that monodisperse polymers exhibit single-exponential relaxation.\cite{zinn_equilibrium_2011} Lu et al. further showed that a logarithmic decay is recovered in solutions with a bimodal chain length distribution.\cite{lu_molecular_2012}

Analyzing exchange kinetics from the decay of scattering intensity requires both a functional form of the escape rate constant, $k_\text{esc}(N_\text{core})$, and the chain length distribution, $f(N_\text{core})$. The dynamic scattering intensity is then expressed as a convolution,
\begin{equation}\label{eq:convolution}
    R(t)=\int_1^\infty dN_\text{core}f(N_\text{core})\exp\left[-k_\text{esc}(N_\text{core})t\right]
\end{equation}
The functional form of the rate constant proposed by Halperin and Alexander can be generalized to include two free parameters, $\alpha$ and $\beta$:\cite{choi_mechanism_2010,lund_equilibrium_2011,lund_equilibrium_2006}
\begin{equation}\label{eq:rate-const}
    k_\text{esc}\sim\frac{1}{\tau_0}\exp\left(-\alpha\chi N_\text{core}^\beta\right)
\end{equation}
where $\alpha$ and $\beta$ depend on the chain conformation and escape mechanism. The exponent $\beta$ ranges from $2/3$ for the Halperin--Alexander collapsed mechanism to $1$ for a core fully exposed to solvent. The prefactor $\tau_0$ sets the timescale and is taken to be the Rouse time, $\tau_0=\tau_R=\xi N_\text{core}^2 l_B^2/(6\pi^2 k_B T)$.\cite{choi_mechanism_2010,lund_equilibrium_2011} The parameter $\chi$ is the monomer-level Flory--Huggins interaction parameter, replacing the macroscopic interfacial tension. The chain length distribution originally used by Lund and coworkers was a Poisson distribution, characteristic of an ideal living anionic polymerization.\cite{lund_equilibrium_2006} Choi et al. later opted for the more flexible Schulz--Zimm distribution that describes imperfect polymerization and can match any experimentally obtained chain length distribution.\cite{choi_mechanism_2010} With this framework, several TR-SANS studies on different polymers and solvents have been used to extract the unimer exchange rate and its dependence on polymer and solvent properties.\cite{choi_mechanism_2010,ma_chain_2016,lund_equilibrium_2011,zinn_equilibrium_2011} In all cases, the free energy barrier scaled linearly with core block length, in direct disagreement with the Halperin--Alexander prediction of $N_\text{core}^{2/3}$ but consistent with theoretical predictions for self-diffusion of BCPs in melts.\cite{cavicchi_self-diffusion_2003,yokoyama_self-diffusion_1998,yokoyama_diffusion_2000}

This discrepancy between the Halperin--Alexander theory and experimental observations prompted extensive discussion and several simulation studies aimed at verifying the linear scaling and elucidating the true escape mechanism. Some studies attempted to replicate the experimental procedure \textit{in silico} by constructing micellar solutions, artificially labeling cores, and monitoring exchange over the course of long unbiased simulations.\cite{li_kinetics_2010,li_equilibrium_2011,prhashanna_kinetics_2016,prhashanna_chain_2017,prhashanna_micelle_2020} While these simulations supported linear scaling with $N_\text{core}$, they did not provide a detailed mechanism for chain escape under experimentally relevant conditions. Namely, \textit{in silico} exchange experiments are required to operate at low enough segregation strength ($\chi$) where a significant number of exchange events can feasibly be observed within the simulation timescale. This is in contrast to experiments where the segregation strength is generally high enough to halt exchange at room temperature on timescales of seconds to hours. In addition, these simulations were limited to core blocks containing only a very small number of coarse-grained beads where the Halperin--Alexander theory would not apply due to the lack of a coil--globule transition.

To resolve these issues, Seeger and coworkers used a different approach relying on enhanced sampling molecular dynamics.\cite{seeger_free_2021,seeger_mechanism_2022} Specifically, they utilized umbrella sampling with the weighted histogram analysis method (WHAM) to compute the free energy profile, or potential of mean force (PMF), of a single chain to escape from an isolated micelle. A similar approach has been used to study the escape free energy of short surfactant molecules.\cite{yuan_potentials_2015,wen_surfactant_2021} With BCPs, this approach allowed them to resolve large free energy barriers for high $\chi$ values and for larger $N_\text{core}$ within a feasible simulation time. They computed a linear scaling of the free energy barrier with $N_\text{core}$ and explained its origin through a simple scaling theory where they assumed the chain escapes "bead-by-bead". Their calculations shed light on a hyperstretching (or "bead-by-bead") escape mechanism as an explanation for the failure of the Halperin--Alexander mechanism to match experimental observations. The term hyperstretching refers to the chain extending far beyond its ideal end-to-end distance. However, these simulations were limited by the use of a single collective variable and still relatively short chain lengths, with $N_\text{core}$ ranging from 4 to 12. Due to the use of a single distance collective variable, they observe a discontinuous jump in the polymer conformation along their effective reaction coordinate. This indicates that there is an additional barrier in the polymer conformation that can lead to incomplete sampling for each value of the chosen distance CV, especially near the transition state.\cite{barducci_metadynamics_2011}

In this work, we address some of the challenges encountered in previous simulation studies by taking two different but complementary approaches. In doing so, we provide a complete picture of the exchange mechanism in the high segregation regime where exchange is rare, both in the case of core collapse and only partial shrinking. First, we utilize coarse-grained molecular dynamics (CGMD) simulations with force-bias enhanced sampling to compute the 2-dimensional free energy surface (FES) of the chain exchange process, where one dimension corresponds to the distance of the chain from the micelle, and the other to the degree of chain extension. Through the use of two collective variables, we can achieve more complete sampling of the chain conformation during the escape process. In agreement with the previous work by Seeger and coworkers, we identify distinct collapsed and extended conformations. As expected, we observe a barrier between the collapsed and extended conformation at the transition state, which highlights the need for external biasing in two collective variables over just one. With the 2d FES, we compute the minimum free energy pathway (MFEP) using the string method and show that it corresponds to the Halperin--Alexander mechanism. We also identify a low free energy region of the FES corresponding to a possible extended escape mechanism that may be kinetically favored under some circumstances in the presence of fluctuations. It is still unclear if there is a regime where the chain is collapsed in the solution and follows a hyperstretching mechanism that would lead to a linear scaling as observed in many experiments.

Additionally, we study the escape mechanism in the high density (polymer melt) limit where the core block does not fully collapse, leaving most or all of the core beads exposed to the unfavorable surroundings. In this regime we employ forward flux sampling (FFS) on a simplified single-chain model that mimics a polymer immersed in a dense melt. In our case, specifically a phase separated copolymer melt. FFS is a transition path sampling technique that introduces no external force biases, and therefore preserves dynamics.\cite{van_erp_novel_2003,allen_forward_2006,borrero_reaction_2007} We compute both the rate of chain escape and the free energy barrier as a function of core block length, and show that the free energy scales linearly with $N_\text{core}$. Additionally, we analyze different ensembles of chain properties during the escape process by extracting full escape trajectories. These ensembles reveal that the chains prefer to escape by first extending ("bead-by-bead") into the solution, and then shrinking.

In the following sections, we start by describing the MD simulation model, enhanced sampling methods, and analysis of the free energy surface for escape within the Halperin--Alexander regime. We then discuss the single-chain model, forward flux sampling, and the escape mechanism within the melt regime.

\section{Coarse-Grained Molecular Dynamics}
\subsection{Methods}
\subsubsection*{Simulation Model}
We model diblock copolymers in an explicit solvent using  highly coarse-grained molecular dynamics simulations. For simplicity, we assume that all particles have the same effective diameter ($\sigma$). In line with previous works, we borrow the conservative force from the DPD potential to describe the non-bonded repulsion between beads,\cite{groot_dissipative_1997,li_kinetics_2010,li_equilibrium_2011,seeger_free_2021,seeger_mechanism_2022,mysona_simulation_2019}
\begin{equation}\label{DPD}
    \beta U_{nb} (r_{ij}) = \frac{1}{2}\beta\epsilon_{ij}\left(1-\frac{r_{ij}}{\sigma}\right)^2,\;\;\;\;\;\;r_{ij}<\sigma
\end{equation}
where $\beta=1/k_B T$, and $\epsilon_{ij}$ is the repulsion strength between particles $i$ and $j$. We choose the base repulsion between all species to be $\epsilon=25\,k_BT$. Incompatible pairs of species such as the core-forming polymer block and monomeric solvent have an $\epsilon_{ij}=48\,k_B T$. Polymeric beads are bonded together using a harmonic spring potential given by,
\begin{equation}\label{harmonic-bonds}
    \beta U_{b}(r_{ij}) = \frac{1}{2}\kappa(r_{ij}-\sigma)^2
\end{equation}
where $\kappa$ is the spring constant, and we use $\sigma$ as the bond length. We use a value of $\kappa=100 \,k_B T/\sigma^2$ which is in line with previous studies of block-copolymer micelles.\cite{seeger_free_2021,seeger_mechanism_2022} In the system there are $n$ polymers each having $N$ monomers, which are divided into two blocks of length $N_A$ and $N_B=N-N_A$. A is the core-forming block, and B is the corona-forming block. The system also contains $N_S$ monomeric solvent molecules, for a total of $nN+N_S$ monomers in the system. In line with previous studies, we utilize a reduced density of $\rho=3.0\sigma^{-3}$. The maximum chain-length that we study is $N=124$, which has an ideal end-to-end distance of $R_\textrm{e2e}=\sigma\sqrt{N-1}\approx 11.1\sigma$. We utilize a box size of $L=55\sigma$ which corresponds to $L\approx 5R_\textrm{e2e}$ to ensure that there are no finite size effects. For the smallest polymer we study, $N_A=4$ yielding an invariant degree of polymerization of $\sqrt{\bar{\mathcal{N}}}=\rho\sigma^3\sqrt{N_A}\approx 6$. The largest polymer we study has $N_A=100$, yielding $\sqrt{\bar{\mathcal{N}}}\approx 30$.

In all simulations, we use $n=36$ chains to form the isolated micelle such that our results are directly comparable to previous works.\cite{seeger_free_2021,seeger_mechanism_2022} Note that the equilibrium micelle size distribution is very wide, and thus there are a large number of reasonable choices for $n$. One has to ensure that $n$ is not so far above the optimal aggregation number such that the micelle undergoes spontaneous fission during the course of a long simulation. The choice of a relatively small $n$ results in a diffuse corona to avoid any enhancement of the exchange rate due to corona crowding across all values of $N_A$ used.\cite{lodge_dynamics_2022} Also note that in the strongly segregated regime, the exchange rate is exceedingly low, such that we do not observe any exchange events that are not a direct result of our biasing methods described below.

\begin{figure}[h!]
    \centering
    \includegraphics[width=0.6\linewidth]{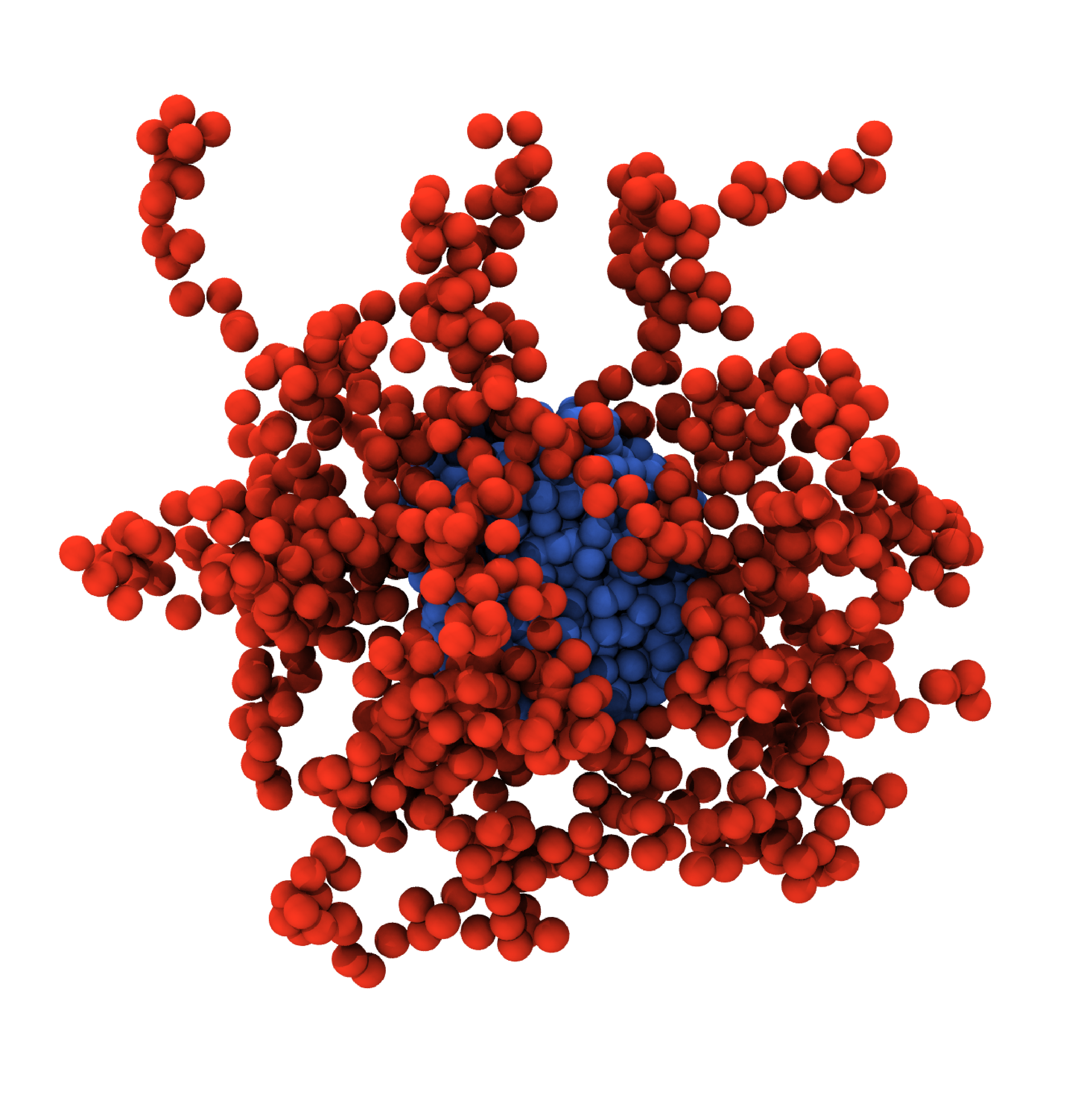}
    \vspace{-0.7cm}
    \caption{Example of a stable micelle with $n=36$, $N_A=21$, $N_B=24$, $\epsilon=\epsilon_{ii}=\epsilon_{BS}=25\,k_BT$, and $\Delta\epsilon=\epsilon_{AS}-\epsilon=\epsilon_{AB}-\epsilon=23\,k_BT$. Solvent particles are omitted for visual clarity.}
    \label{fig:example-micelle}
\end{figure}

We run our simulations in OpenMM\cite{eastman_openmm_2023} by making use of the open-source MDCraft\cite{ye_mdcraft_2024} python package that contains helper functions and additional custom non-bonded potentials. We use the \textit{middle} Langevin integrator with a time-step of $\Delta t = 0.01\tau$ and a friction coefficient of $\eta=1/\tau$, where $\tau=\sqrt{m\sigma^2/(k_B T)}$. See Figure \ref{fig:example-micelle} below for a visual example of a stable micelle.

\subsubsection*{Enhanced Sampling}
In order to compute the free energy barrier for chain expulsion, we employ enhanced sampling to bias collective variables (CVs) between low and high free energy regions of the phase space. To define our CVs, we separate the type A atoms into two groups: (1) $N_A(n-1)$ atoms forming the core of the micelle, which includes all chains minus one, and (2) the $N_A$ atoms of a selected chain which will undergo expulsion. We define the coordinates of the atoms in group 1 as $\*R$, and the atoms of group 2 as $\*r$. We define Basin1 as the stable basin in which the selected chain is within the micelle, and Basin2 as the metastable plateau region in which the selected chain has escaped and no longer \textit{sees} the micelle. We define two different distance-based CVs to track the progress of the system between Basin1 and Basin2. The first is the distance between the center of mass of the micelle (excluding the selected chain) and the junction point of the selected chain, where the junction refers to the point of connection between the A and B blocks.
\begin{equation}
    \text{CV1} = R_\textrm{cm-jp} = || \*R_\textrm{cm} - \*r_\textrm{jp} ||_2
\end{equation}
The second is the distance between the center of mass of the micelle (excluding the selected chain) and the center of mass of the core block of the selected chain.
\begin{equation}
    \text{CV1}^\prime = R_\textrm{cm-cm} = || \*R_\textrm{cm} - \*r_\textrm{cm} ||_2
\end{equation}
Previous studies have utilized $R_\textrm{cm-jp}$ to conduct umbrella sampling simulations,\cite{seeger_free_2021,seeger_mechanism_2022} where free energy profiles are constructed using the weighted histogram analysis method (WHAM). However, we expect that a single CV is not sufficient to obtain an accurate free energy estimate due to the possible presence of barriers in other collective variables. The presence of barriers in orthogonal CVs causes insufficient sampling in configurational space.\cite{barducci_metadynamics_2011} Namely, in this case, the polymer conformation can range from fully extended to fully collapsed, however, this full spectrum cannot be readily explored at each value of $R_\textrm{cm-jp}$ due to significant barriers in changing the polymer conformation. To remedy this, we propose running simulations with two collective variables simultaneously, which has become much more feasible in recent years due to advancements in enhanced sampling methods and accessibility of high performance graphical processing units (GPUs). We define a third collective variable, $r_\textrm{e2e}$ to be used in conjunction with either of the two distance based CVs defined above. $r_\textrm{e2e}$ is the end-to-end distance of the core block (A block) of the selected chain.
\begin{equation}
    \text{CV2} = r_\textrm{e2e} = || \*r_{N_A} - \*r_1 ||_2
\end{equation}
This collective variable allows us to bias the conformation of the escaping chain to sample the full range from fully collapsed to fully extended. To clarify, we denote the overall end-to-end distance of the chain as $R_\textrm{e2e}$ and the end-to-end distance of the core block only as $r_\textrm{e2e}$.

We compute the 2d FES for the combinations $\{R_\textrm{cm-jp}$,\,$r_\textrm{e2e}\}$ and $\{R_\textrm{cm-cm}$,\,$r_\textrm{e2e}\}$ for various different values of $N_A$ and $\Delta \epsilon$ to elucidate the preferred escape mechanism and the scaling relationships of the free energy barrier (exchange rate). We study both combinations of CVs to ensure that the results are independent of the choice of CV. We utilize the recently developed Spectral Adaptive Biasing Force (SABF) method available in the PySAGES enhanced sampling package.\cite{zubieta_rico_pysages_2024,zubieta_rico_efficient_2025} SABF is an improved version of the ABF method, that has improved efficiency and stability. ABF-type methods also have an advantage over metadynamics-type methods for our particular system because our CVs all have hard boundaries at 0, which poses a problem for metadynamics but not for ABF-type methods.\cite{mcgovern_boundary_2013}

\subsection{Results and Discussion}

\begin{figure*}[ht!]
    \centering
    \includegraphics[width=1\linewidth]{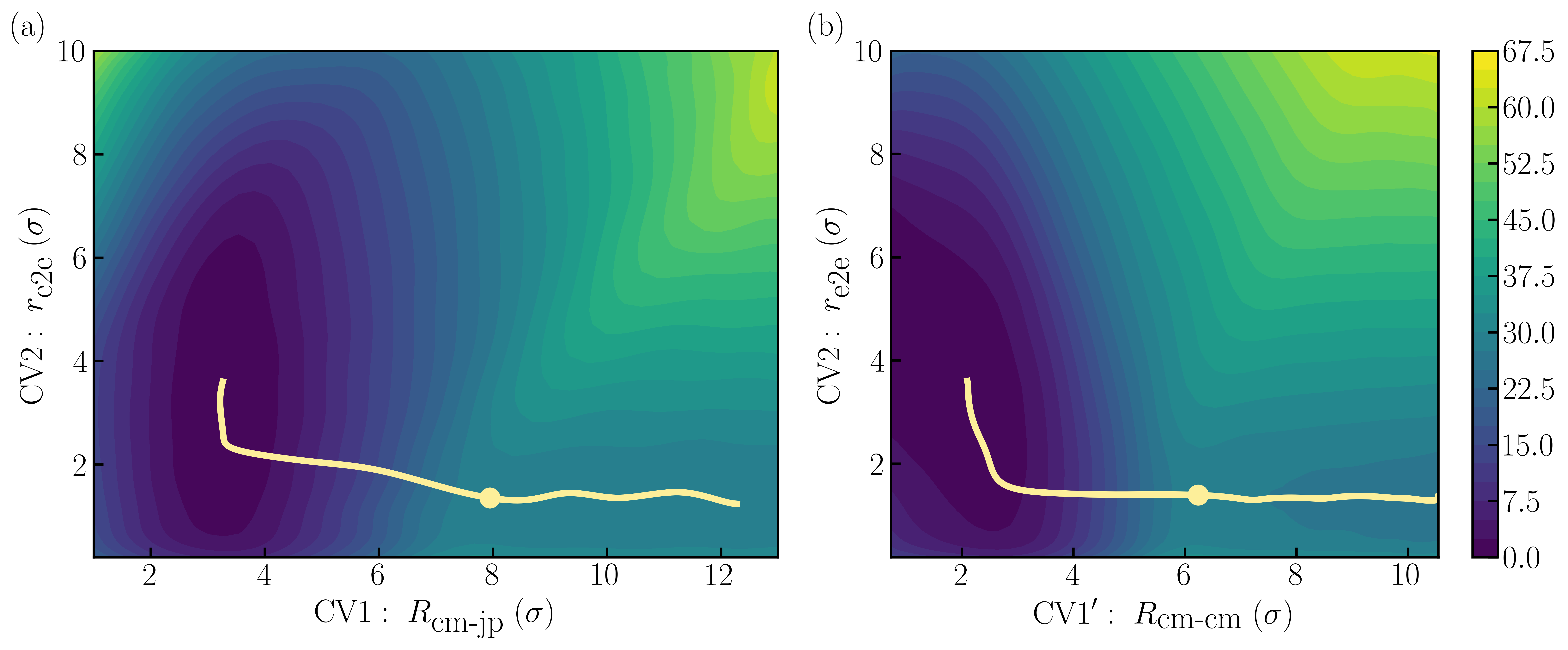}
    \vspace{-0.8cm}
    \caption{2-dimensional FES for CV pairs $R_\textrm{cm-jp}$,$r_\textrm{e2e}$ (a) and $R_\textrm{cm-cm}$,$r_\textrm{e2e}$ (b) with $N_A=15$, $N_B=24$, $\beta\Delta\epsilon=23$. Each surface is shifted such that the minimum free energy is 0. The red lines trace the MFEPs as computed by the string method, with the yellow circles indicating the transition states.}
    \label{fig:twoFES}
\end{figure*}

First, we compute the 2-dimensional FES for both pairs of CVs, $\{R_\textrm{cm-jp}$,$r_\textrm{e2e}\}$ and $\{R_\textrm{cm-cm}$,$r_\textrm{e2e}\}$ for the particular case of $N_A=15,N_B=24$ and $\beta\Delta\epsilon=23$. Both surfaces are presented in Figure \ref{fig:twoFES}. In the following discussion, we refer to the surfaces in Figures \ref{fig:twoFES}a and \ref{fig:twoFES}b as FES1 and FES2, respectively. Both FES1 and FES2 have a significant free energy basin at low values of all CVs, which corresponds to the selected chain being located within the micelle. Note that when the chain is within the micelle, $r_\textrm{e2e}$ can take on a wide range of values at low free energy cost. Large fluctuations in $r_\textrm{e2e}$ are expected since the micelle core presents a \textit{theta} solvent environment. The average value of the core block end-to-end distance in the basin is $\langle r_\textrm{e2e} \rangle = 3.3$ which is reasonably close to the ideal value of $\sqrt{N_A-1}=3.7$ for a freely-jointed chain. FES1 displays a near-vertical basin, indicating that the chain may extend and collapse while the junction point is consistently localized to the surface of the micelle. From FES1, we compute the average value of the relative position of the junction point in the basin to be $\langle R_\textrm{cm-jp}\rangle \approx 3.4$. This agrees with the radius of the micelle, as shown in Figure S1 in the ESI$^\dagger$.

FES1 (FES2) is characterized by the presence of a large basin for low $R_\textrm{cm-jp}$ ($R_\textrm{cm-cm}$) and a plateau for high values of $R_\textrm{cm-jp}$ ($R_\textrm{cm-cm}$). The plateau at high $R_\textrm{cm-jp}$ ($R_\textrm{cm-cm}$) and low $r_\textrm{e2e}$ corresponds to an escaped chain that is collapsed in solution. It is clear from both FES1 and FES2 that there exists a pathway where the chain exits the micelle in a collapsed state. This is indicated by the entrance to the plateau (tube) being centered around $r_\textrm{e2e}\approx 1$. This pathway is qualitatively consistent with the collapsed Halperin--Alexander mechanism.\cite{halperin_polymeric_1989,halperin_micellar_2011}

While the basin and plateau are the two main features of FES2, FES1 has additional interesting behavior at intermediate values of $R_\textrm{cm-jp}$ and $r_\textrm{e2e}$. In this region the chain has partially escaped, but is able to take on an extended conformation with some monomers still located within the micelle core. For $R_\textrm{cm-jp}$ in the range of (5,8), the extended conformation has a lower free energy than the collapsed conformation. This becomes more obvious when we plot the conditional probability distribution, $P(r_\textrm{e2e}|R_\textrm{cm-jp})$, which is calculated directly from the FES using Equations \eqref{prob} and \eqref{condprob}.
\begin{equation}\label{prob}
    P(r_\textrm{e2e},R_\textrm{cm-jp})=\exp\left[-\beta\Delta F(R_\textrm{cm-jp},r_\textrm{e2e})\right]
\end{equation}
\begin{equation}\label{condprob}
    P(r_\textrm{e2e}|R_\textrm{cm-jp})=\frac{P(r_\textrm{e2e},R_\textrm{cm-jp})}{\int \mathrm{d}r_\textrm{e2e} P(r_\textrm{e2e},R_\textrm{cm-jp})}
\end{equation}

\begin{figure}[b!]
    \centering
    \includegraphics[width=1.0\linewidth]{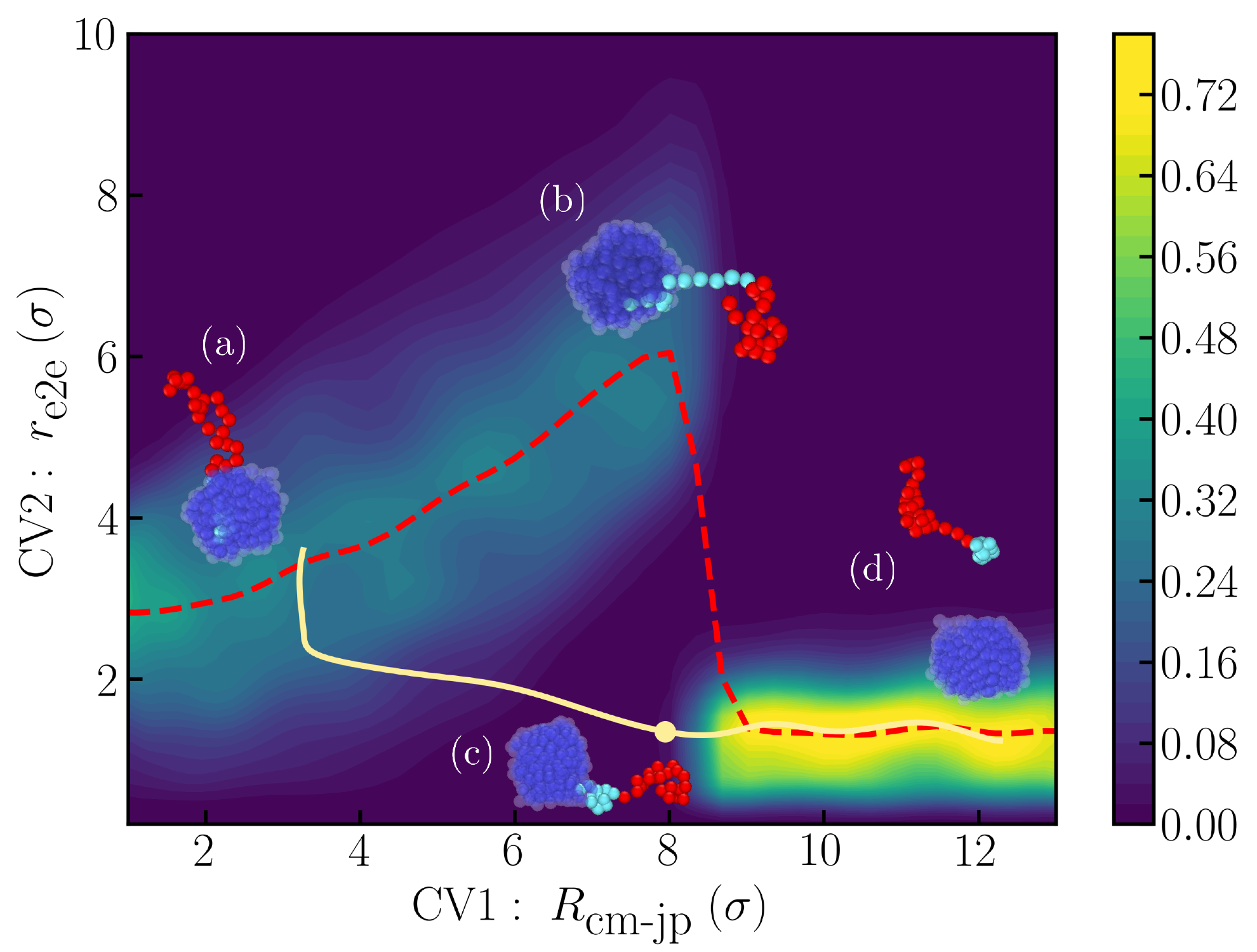}
    \vspace{-0.8cm}
    \caption{Conditional probability distribution for the core block end-to-end distance as a function of the location of the junction point for $N_A=15$, $N_B=24$, $\beta\Delta\epsilon=23$. The distribution is computed using FES1 from Figure \ref{fig:twoFES}a. The red-dashed line traces the mean of $r_\textrm{e2e}$ from the conditional distribution, and the yellow line traces the MFEP from FES1. Visualizations are (a) the chain in the micelle, (b) the chain extended into solution, (c) the chain collapsed at the micelle interface and (d) the chain fully expelled.}
    \label{fig:condProb}
\end{figure}
The resulting distribution shown in Figure \ref{fig:condProb} indicates that an extended conformation is actually more probable prior to complete escape. At the saddle point position of CV1 (red dot), we find that the extended conformation (large CV2) has a lower free energy, indicated by the higher conditional probability density. This suggests that the chain may escape first by extending into solution until the contact with solvent is too unfavorable, at which point the chain collapses, expelling the remaining beads and forming a compact globule. This analysis assumes that the chain has enough time to fully relax at each value of $R_\textrm{cm-jp}$ during the expulsion process. These results agree with and further support previous findings by Seeger et al. who used umbrella sampling simulations to compute 1-dimensional potentials of mean force (PMFs) for chain expulsion.\cite{seeger_free_2021,seeger_mechanism_2022} They found that $R_g$ gradually increased with $R_\textrm{cm-jp}$ up to a certain point, where the chain then collapsed. Similarly, we also observe a bimodal distribution in $P(r_\textrm{e2e}|R_\textrm{cm-jp})$ near the transition state.

\begin{figure}[ht!]
    \centering
    \includegraphics[width=1.0\linewidth]{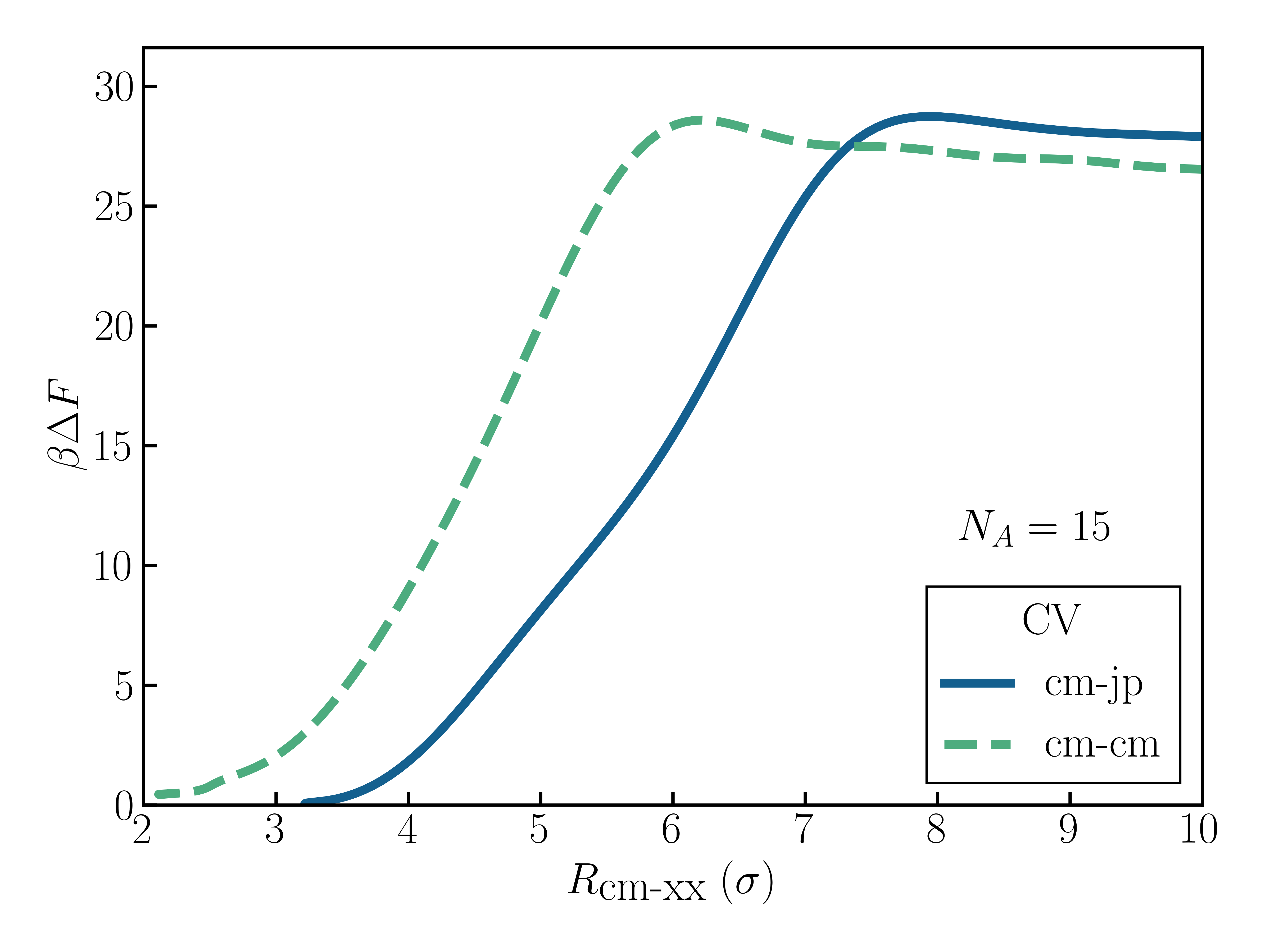}
    \vspace{-0.8cm}
    \caption{The MFEP plotted along a single dimension, $R_\textrm{cm-xx}$. The MFEP are extracted from the 2d FES in Figure \ref{fig:twoFES}.}
    \label{fig:twoMFEP1d}
\end{figure}

In addition to the FES and conditional distribution, it is of interest to compute the 1-dimensional free energy profile along an effective reaction coordinate. For this we consider both the minimum free energy path (MFEP) and free energy projection. First, we compute the MFEP via the string method (see the ESI$^\dagger$ for method details).\cite{e_string_2002,e_simplified_2007,e_transition-path_2010} We conduct the string method optimization on the already computed 2d FES; we do not employ the string method during the MD simulations themselves. In Figure \ref{fig:twoFES}, we plot the MFEP on top of both FES1 and FES2. We also plot the MFEP in one dimension as a function of only the $R_\textrm{cm-xx}$ collective variables in Figure \ref{fig:twoMFEP1d}. Note that the two MFEP are nearly identical barring a horizontal shift.
\begin{figure*}[ht!]
    \centering
    \includegraphics[width=1.0\linewidth]{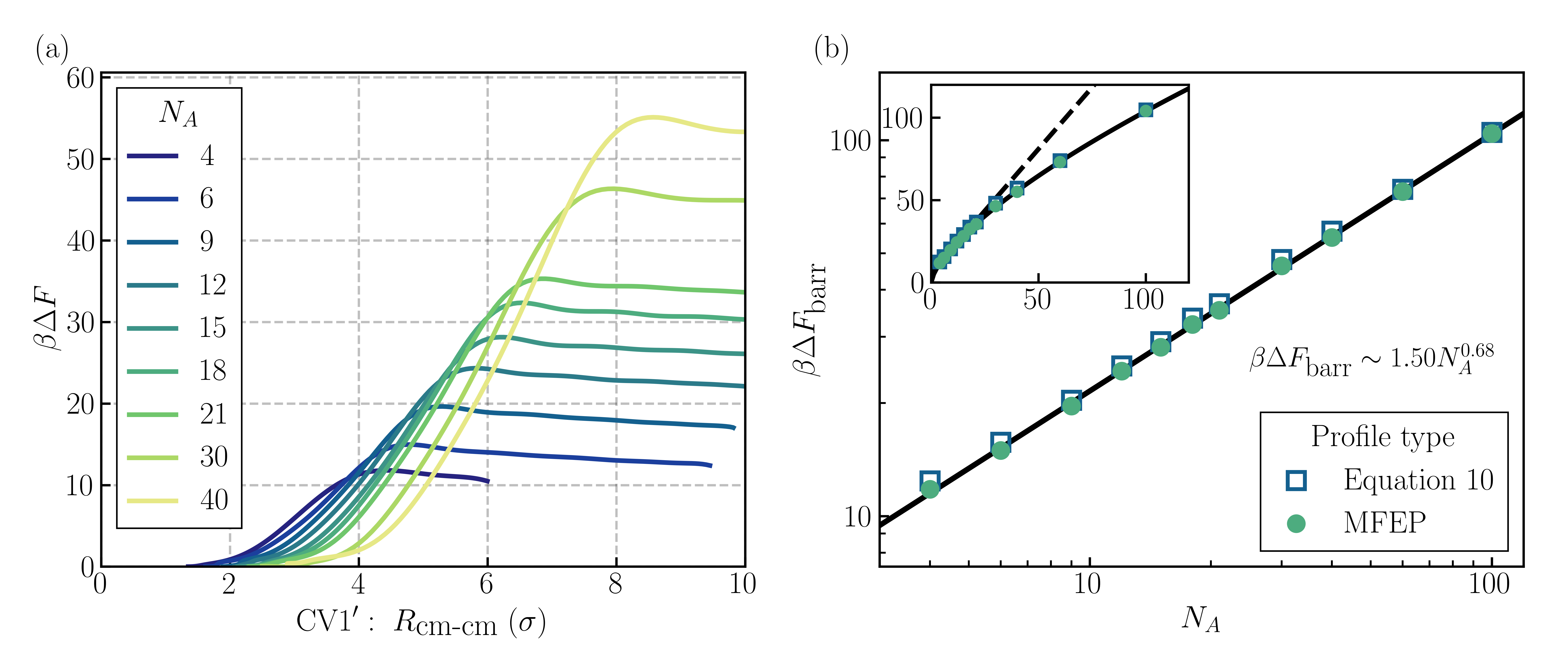}
    \vspace{-0.8cm}
    \caption{(a) 1-dimensional free energy profile of the MFEP for various core block lengths, $N_A$. Note that all MFEP are computed on a 2d FES similar to FES2 from Figure \ref{fig:twoFES}. (b) Free energy barriers from MFEP and free energy projection plotted against core block length, $N_A$, on a log-log scale. The solid line regression was conducted for the expression, $\ln(\beta\Delta F_\text{barr})=a\ln(N_A)+b$, where $a$ and $b$ were fitting parameters, and the MFEP was used. The inset is the same as (b) on a linear--linear scale, and the dotted line is a linear regression, $\beta \Delta F\text{barr}=aN_A+b$, of the first five points corresponding to the small $N_A$ region.}
    \label{fig:regression}
\end{figure*}

The computed MFEP qualitatively follows the Halperin-Alexander (HA) picture of micelle chain exchange. In the HA mechanism, the chain escapes the micelle in a collapsed state, resulting in a free energy barrier that scales with $N_A^{2/3}$. In order to confirm that the MFEP computed from our simulations yields the same scaling as the HA theory, we computed the MFEP for a range of core block lengths and computed the scaling relationship. The free energy curves and resulting regression analysis are given in Figure \ref{fig:regression}. In Figure \ref{fig:regression}b we find that the scaling is very near $2/3$, providing further support that the MFEP follows the HA mechanism, and that a $2/3$ scaling does exist under conditions where core collapse is expected. In the inset of Figure \ref{fig:regression}b we plot the free energy barrier as a function of the core block length on a linear--linear scale. Interestingly, we find that a linear fit is reasonable at small $N_A$, consistent with the results of previous simulations,\cite{li_equilibrium_2011,seeger_mechanism_2022} and with the expectation that very short core blocks cannot collapse to effectively shield monomers from the solvent.

In addition to the MFEP, we project the 2-dimensional free energy surface into 1 dimension corresponding to the $R_\textrm{cm-cm}$ pseudo-reaction coordinate. This analysis allows us to draw a more direct comparison with the simulations of Seeger et al. where only a single collective variable was biased.\cite{seeger_free_2021,seeger_mechanism_2022} In principle, the projection should be more reliable than the direct single collective variable calculation as the dual collective variable simulation enables much more complete sampling of the polymer conformation ($r_\textrm{e2e}$). The projection is done using the normalization constant in Equation \eqref{condprob} which accumulates the weight of the free energy surface at each $R_\textrm{cm-cm}$.
\begin{equation}\label{projectedF}
    \beta F(R_\textrm{cm-cm})=-\ln\left[\int dr_\textrm{e2e} P(R_\textrm{cm-cm},r_\textrm{e2e})\right]
\end{equation}
We plot the barriers of the 1-dimensional free energy profiles as blue squares in Figure \ref{fig:regression}b. We find that the free energy barriers computed from the projected free energy are only slightly higher than those computed from the MFEP, and therefore still yield the same $N_A^{2/3}$ scaling of the free energy barrier, in contrast to the linear scaling obtained in previous simulations.\cite{seeger_mechanism_2022} We provide a direct comparison of the barriers obtained from the two different methods in the ESI$^\dagger$.

We can foresee two potential reasons for the discrepancy, with the first being simple and the latter being rather complicated. The simple explanation is that a linear scaling should be observed at short chain lengths due to incomplete collapse of the core block, as shown in the inset of Figure \ref{fig:regression}b. While this is true for very short chain lengths, we still found a significant deviation from a linear scaling beyond $N_A=8$, while Seeger et al. studied chains up to $N_A=12$ and still found linear scaling.  The other, more probable reason, is the presence of the ridge between the collapsed and extended conformations on FES1 of Figure \ref{fig:regression}a. This ridge is consistent with the observations made by Seeger et al. that the $R_g$ of the core block becomes bimodal as $R_\textrm{cm-jp}$ approaches the transition state. In simulations with a single CV, one could encounter hysteresis that can affect the computed barrier. If the chain starts in the micelle and is progressively pulled out, then the upper valley on FES1 will be preferred since the free energy gradient there is initially lower. The chain can remain in the extended conformation beyond the saddle point value of CV1 in FES1 (yellow dot) due to the significant barrier between the two conformations which would lead to a delayed transition state and an enhanced free energy barrier. The ridge between the two conformations increases in height for larger values of $N_A$ which can impact the scaling of the barrier with $N_A$. If the chain is instead pushed into the micelle (reverse direction), then we expect that the MFEP will be followed and the observed barrier will be lower. This serves as a reasonable explanation for why the 1d umbrella sampling simulations yielded an apparent linear scaling, whereas our simulations yielded the Halperin--Alexander 2/3 scaling that is expected for large $\chi$ and large $N_A$.

With these results we have qualitatively and quantitatively identified the Halperin--Alexander collapsed mechanism as the MFEP for chain escape under strong segregation at sufficiently long core-block lengths, and have provided further context for the linear scaling observed in DPD simulations of shorter chains.\cite{li_equilibrium_2011,seeger_mechanism_2022} Additionally, we have shown that the extended conformation observed by Seeger et al.\cite{seeger_free_2021,seeger_mechanism_2022} is a valley on the free energy landscape that is actually slightly more favorable than the collapsed chain prior to the transition state. Thus, the chain may attempt to escape more frequently by fluctuating out of the micelle "bead-by-bead", at which point it is met with an additional barrier to collapse and fully escape. Physically, this additional free energy cost is associated with exposing the remaining monomers to the solvent while keeping the junction point fixed. On the other hand, the chain may attempt to escape less frequently by first collapsing within the micelle core, but when it does, it is met by a lower free energy barrier due to having a minimal number of contacts with the solvent.

The question still remains as to why the MFEP from these simulations is at odds with experimental observations in terms of the scaling of the free energy barrier. As mentioned previously, one can expect a crossover from linear scaling of the barrier for a chain fully exposed to the solvent, to the Halperin--Alexander 2/3 scaling for a collapsed chain. The free energy barrier for a solvated chain in DPD is $\beta F_\textrm{solv}\sim \beta\Delta\epsilon N_A / (\rho\sigma^3)$, while the barrier for a collapsed chain is approximately $\beta F_\textrm{HA}\sim \beta\Delta\epsilon N_A^{2/3}(\rho\sigma^3)^{-2/3}$, ignoring constant prefactors. Thus, the ratio of the two barriers is expected to scale as $F_\textrm{solv}/F_\textrm{HA}\sim (N_A/\rho\sigma^3)^{1/3}$. In other words, the scaling should be linear with $N_A$ when the effective coordination number of a monomer with the solvent is much higher than $N_A$. The scaling should go as $N_A^{2/3}$ when the coordination number is much lower than $N_A$, since a large portion of the monomers can replace solvent contacts with other monomer contacts upon collapse. Therefore, at low densities and long chain lengths, the chain should be collapsed and follow the Halperin--Alexander mechanism. In our DPD simulations with $\rho\sigma^3=3$ and $N_A$ up to 100, we are comfortably within the Halperin--Alexander regime. This is further validated in the ESI$^\dagger$ where we plot the average end-to-end distance of the core block and find that it is fully collapsed within the solvent for most of the studied chain lengths.

In experiments, it is unclear and highly situational whether the system corresponds more to the polymer melt case with only a partially shrunken core block upon escape, or the Halperin--Alexander case with a dry collapsed core block. For hairy micelles with very dilute coronas (as in our simulations), it is expected that the core block should escape in a collapsed state and produce $N_A^{2/3}$ scaling. Indeed, Lund et al. measured chain exchange in micelles formed from highly asymmetric PEP1-PEO20 and found that the exchange barrier could be fit well with a 2/3 power law.\cite{lund_equilibrium_2006} For crew cut micelles that have a dense corona, the escape of the core could be viewed as escaping into the corona domain, rather than directly into the solvent. If the corona is sufficiently dense, Lund et al. argued that the increased pressure could prevent the core from collapsing, and lead to a barrier scaling linearly with $N_A$.\cite{lund_equilibrium_2011} This was corroborated by exchange measurements they conducted on symmetric PEP1-PEO1 where they computed a linear exchange barrier. They use a simple blob scaling argument to determine when the density of the corona is high enough to prevent collapse of the core block upon escape and verified that their prediction was consistent with their PEP1-PEO20 and PEP1-PEO1 systems as well as the PS-PEP/squalene system of Choi, Lodge, and Bates.\cite{choi_mechanism_2010} In summary, whether the free energy barrier will scale as $N_A$ or $N_A^{2/3}$ depends directly on the ability of the core block to collapse in the unfavorable domain. Our simulation results clearly show that the DPD model and chain/micelle parameters used here correspond to the Halperin--Alexander case.


We can also call into question the underlying assumptions of the zero-temperature string method and the nature of the MFEP. Namely, the computed MFEP only considers the structure of the underlying FES, and ignores any effects of thermal fluctuations or chain dynamics. As a result, the MFEP is most reliable when it corresponds to a deep valley or saddle on the free energy landscape. In addition, the kinetic pathway will only mimic the MFEP if the duration of an escape trajectory is significantly longer than the chain relaxation time, or the time. Our computed 2d FES does not feature a deep transition \textit{tube}, but rather two possible competing free-energy valleys connected by a continuous distribution of pathways with nearly degenerate free energies.

In the following section, we study the other dominant regime corresponding to a very dense polymer melt, where the core block shrinks upon escaping, rather than collapsing into a compact globule. As discussed, this could be representative of a micelle with a dense corona, as in the crew cut case. To study the chain escape at high density, we employ a single-chain model that is appropriate for high $\bar{\mathcal{N}}$ systems wherein the interaction of the tagged chain with other chains can be accurately represented instead by interactions with a mean-field background.\cite{muller_dynamics_2021} To avoid the assumptions of the MFEP, we turned to an alternative method that would allow us to determine the kinetic pathway traversed by escaping chains. This is preferable to the MFEP in our case since we expect that fluctuations could play a significant role in how the chain explores the free energy landscape. We utilize a transition path sampling method known as Forward Flux Sampling that can resolve the ensemble of escape trajectories, including one or both of the mechanisms implied from the FES obtained from our MD simulations.

\begin{figure}[b!]
    \centering
    \includegraphics[width=1.0\linewidth]{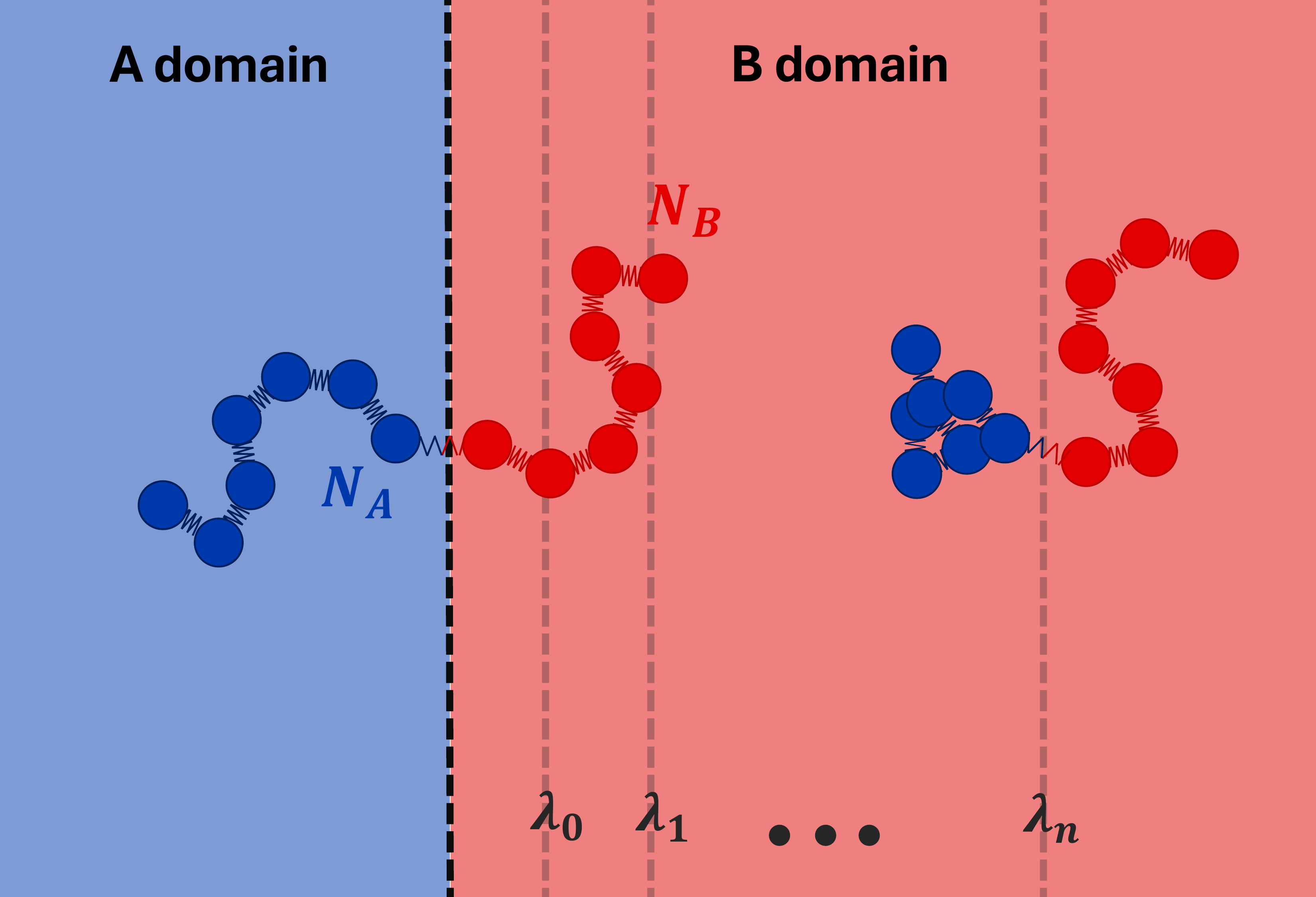}
    \vspace{-0.8cm}
    \caption{A visual representation of the MC simulation setup. The two bead-spring chains represent the same chain at different points in the expulsion process. The left chain is localized to the interface, and the right chain has escaped and collapsed. The dashed lines represent virtual interfaces used for FFS, and they are placed at chosen values of the reaction coordinate (order parameter) $\lambda$.}
    \label{fig:mc-setup}
\end{figure}

\section{Single-Chain Simulations}

In order to study the kinetic pathway for chain escape from a dense micellar core environment, we apply a simpler surrogate model. We do this because the essence of the problem is not the micellar structure itself, but rather the escape of a core from its own melt-like environment. The spherical geometry is not important, nor is the fact that the  field felt by the escaping block is provided by other chains. Thus, we study a single bead-spring polymer immersed in a mean-field background, escaping from a planar interface between two polymer melts. The model is similar to the model used previously by Helfand to study diffusion in strongly segregated copolymer melts,\cite{helfand_diffusion_1992} and by M\"{u}ller to study bridge-loop conversion in lamellae-forming triblock copolymers.\cite{muller_dynamics_2021} We run Metropolis Monte Carlo simulations coupled with Forward Flux Sampling (FFS) to compute the escape rate and observe the unbiased escape trajectories. As a result, we are able to verify the scaling of the exchange barrier as well as the qualitative mechanism for chain escape.

\subsection{Methods}
\subsubsection*{Simulation Model}

We utilize a soft particle simulation model that is similar to that used previously to study dynamic single-chain processes in dense polymer systems.\cite{muller_studying_2011,daoulas_single_2006,schneider_multi-architecture_2019,muller_dynamics_2021} We consider a single polymer chain with bead coordinates $\*r_i$ for $i\in 0,1,\cdots,N-1$. The polymer beads are connected by harmonic springs with the following potential,
\begin{equation}
    \frac{H_b}{k_B T} = \sum_{i=1}^{N-1}\frac{3}{2\sigma^2}\left|r_i-r_{i-1}\right|^2 = \sum_{i=1}^{N-1}\frac{3(N-1)}{2R_\textrm{e2e}^2}\left|r_i-r_{i-1}\right|^2
\end{equation}
where $R_\textrm{e2e}$ denotes the ideal root-mean-square end-to-end distance, $R_\textrm{e2e}^2=(N-1)\sigma^2$, and $\sigma$ is the statistical segment length.  We divide the chain into two blocks, with the first $N_A$ beads belonging to block A and the final $N_B=N-N_A$ beads belonging to block B. The non-bonded interactions consist of two contributions, $H_\text{nb}=H_\text{ext}+H_\text{pair}$. The first term represents the interactions between the polymer beads and the surrounding background fluid, which is a melt of diblock copolymers of the same nature. The background is static and gives rise to effective fields, such that the Hamiltonian can be written as
\begin{equation}\label{nb}
    \frac{H_\text{ext}}{k_B T} = \sum_{i=0}^{N_A-1}w_A(\*r_i)+\sum_{i=N_A}^{N-1} w_B(\*r_i)=\sum_{i=0}^{N-1}w_{t(i)}(\*r_i)
\end{equation}
where $t(i)$ is the type of bead $i$, either A or B. The fields, $w_A(\*r)$ and $w_B(\*r)$ are parameters of the model and are not impacted by the presence of the tagged polymer. Conceptually, these interactions represent the interactions of a given polymer bead with the beads in the surrounding environment, where unfavorable AB contacts increase the energy by $\epsilon k_B T$, while AA and BB contacts decrease the energy by the same amount. Let $z_c$ denote the average number of contacts of a single polymer bead (including inter- and intramolecular contacts). If the composition of the background medium is denoted as $\phi_A$ and $\phi_B=1-\phi_A$, then the interaction strength between the chain and the background can be approximated as $w_A=-z_c\epsilon(\phi_A-\phi_B)$. Similarly, we have $w_B=-z_c\epsilon(\phi_B-\phi_A)$. Note that in an A-rich domain, the energy of an A segment is $-z_c\epsilon$, while the energy of a B segment is $z_c\epsilon$. This energy can be mapped to the Flory-Huggins model with $\chi\approx 2\epsilon z_c$, since that is the difference in energy for an A segment to go from an A-rich domain to a B-rich domain.

\begin{figure*}[hbt!]
    \centering
    \includegraphics[width=1.0\linewidth]{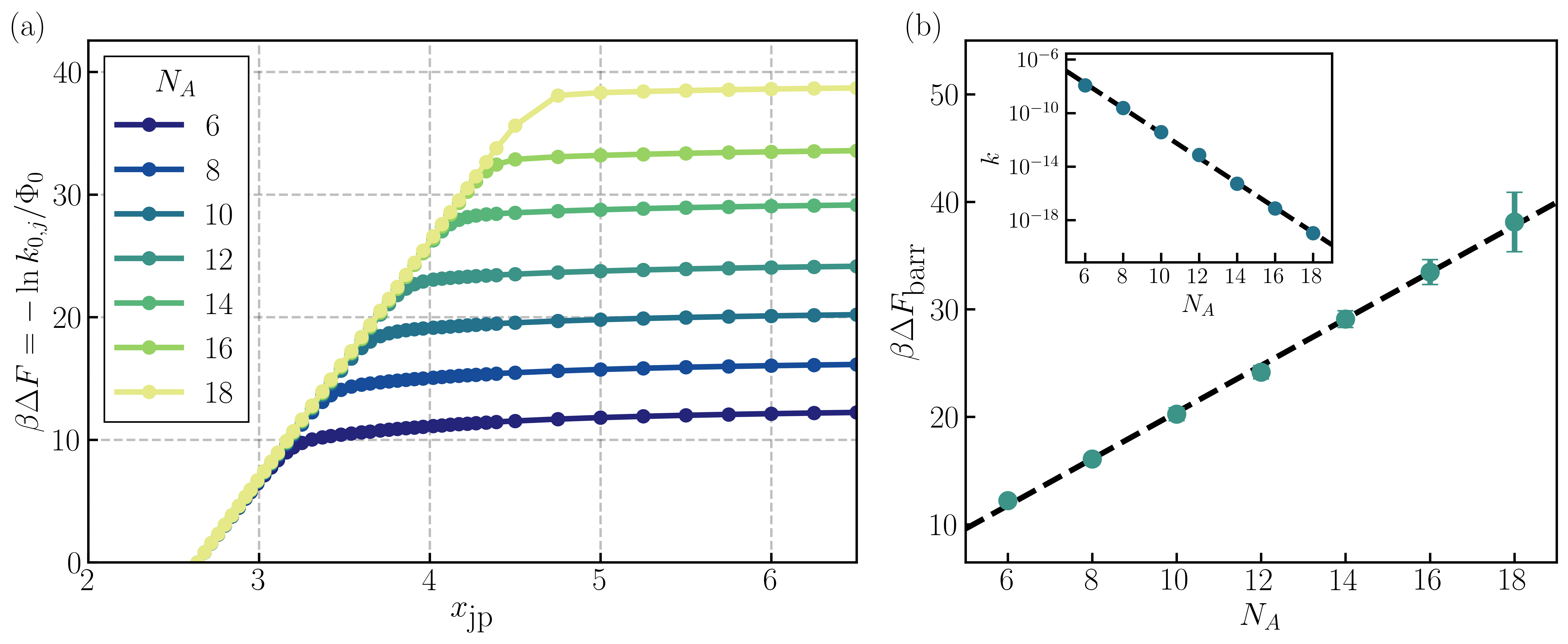}
    \vspace{-0.8cm}
    \caption{FFS results for the $x_\textrm{jp}$ generating CV with various core-block lengths, $N_A$. All lengths are scaled by $R_\textrm{e2e}$, with the interface placed at $x=2.5$. In all cases, the total chain length is $N=32$. (a) The cumulative free energy change. (b) The total free energy change from Equation \eqref{eq:ffs-fe} versus core-block length, with the dashed line being the optimized linear fit. Error bars represent a 95\% confidence interval from averaging 5 independent FFS simulations. The inset is the rate constant from Equation \eqref{eq:ffs-rate}.}
    \label{fig:jun-rates}
\end{figure*}

Some non-bonded interactions come from intramolecular contacts, and these can be accounted for explicitly in the single-chain Hamiltonian,
\begin{multline}
    \frac{H_\text{pair}}{k_B T} = \sum_{i<j} v(\*r_i-\*r_j)\biggl\{-\epsilon[2\delta(t(i),t(j))-1]-\\[1pt]
    \frac{w_{t(i)}(\*r_i)+w_{t(j)}(\*r_j)}{2z_c}\biggr\}
\end{multline}
where pairwise contacts are defined by $v(\Delta\*r)=1$ for $|\Delta \*r|<b=R_e/\sqrt{N-1}$, and otherwise 0. For simplicity, we set the two microscopic length scales -- statistical segment length and interaction range -- equal. $\delta$ is the Kronecker delta function, which is 1 when the segments are of the same type (i.e. $t(i)=t(j)$), and 0 otherwise. The first term quantifies the energy of an intramolecular contact. The second term corrects for double counting, since Equation \eqref{nb} already includes interactions with $z_c$ contacts. We need to subtract any assumed contacts that can be replaced with true intramolecular contacts from beads within the same chain. The number of intramolecular contacts depends largely on the chain conformation, and therefore also on the surrounding environment. A chain of A beads will be extended in an A-rich domain, yielding a low number of intramolecular contacts. Conversely, the chain will be more collapsed in a B-rich domain, yielding a high number of intramolecular contacts. Thus, this model includes the essential physics underlying the collapsed and extended conformations as the chain crosses the interface.

With a uniform density, the total number of contacts is given by $z_c=\frac{4\pi}{3}b^3\rho-1$, where $\rho$ is the segment density in the multichain system. In the high-density limit, we can obtain an approximate relation between $z_c$ and the invariant degree of polymerization, $\bar{\mathcal{N}}$.
\begin{equation}
    z_c\approx \frac{4\pi}{3}\left(\frac{R_e}{\sqrt{N-1}}\right)^3\rho-1\approx \frac{4\pi}{3}\sqrt{\frac{\bar{\mathcal{N}}}{N}}
\end{equation}

In the following, the polymer contains $N=32$ segments, $\epsilon=0.02$, and $z_c=50$. This corresponds to a system with  $\chi N\approx 64$, and $\sqrt{\bar{\mathcal{N}}}\approx 67.5$. See Figure \ref{fig:mc-setup} for a schematic of the system setup.

\subsubsection*{Monte Carlo Simulation}

We run Metropolis Monte Carlo simulations of $\mathcal{O}(10^6)$ independent chains. One MC step consists of selecting a polymer bead, updating its position via $\*r_{v'} = \*r_{v} + (\sigma/\sqrt{N-1})\pmb{\hat{\mathcal{N}}}(0,1)$, computing the new Hamiltonian, and accepting or rejecting the moved based on the Metropolis criterion. Here, $\pmb{\hat{\mathcal{N}}}(0,1)$ is the standard normal distribution. The order of polymer beads is chosen randomly without replacement. This updating scheme is intended to mimic the Rouse dynamics of a polymer chain in a melt,\cite{muller_dynamics_2021,daoulas_single_2006,schneider_multi-architecture_2019} which is appropriate for diffusion of a polymer chain perpendicular to an interface.\cite{lodge_dynamics_2022,yokoyama_self-diffusion_1998,cavicchi_self-diffusion_2003} A detailed description of the algorithm is provided in the ESI$^\dagger$.

\subsubsection*{Forward-flux Sampling}

We utilize forward flux sampling (FFS)\cite{allen_forward_2006,borrero_reaction_2007,hussain_studying_2020} to compute the rate for chains localized to an interface to fully escape into solution. A detailed description of the FFS algorithm is provided in the ESI$^\dagger$. Briefly, FFS is a transition path sampling (TPS) technique that is used to compute the rate of rare events in a way that introduces no external biasing potential or forces. In FFS, virtual interfaces in collective variable (CV) space are placed at regular intervals between the starting and ending basin of the transition path. Transition trajectories are built up by simulating small transitions from one interface to the next, which are by themselves much more probable than the full transition. A \textit{generating CV} is used to define the location of the virtual interfaces and track the progress of each chain from one interface to the next. Ultimately, the trajectories and transition rates can be accumulated from all of the interfaces to compute the overall rate and the ensemble of completed reaction trajectories.

We choose the inhomogeneous external fields, $w_A$ and $w_B$, such that there is an interface located at $x_I=2.5$ in units of $R_\textrm{e2e}=\sigma\sqrt{N-1}$. For $x<x_I$, the external field mimics an A-rich domain at the mean-field level, while $x>x_I$ mimics a B-rich domain. We define Basin1 to be when the chain is localized to the interface, $x_\textrm{jp}\approx x_I$, where $x_\textrm{jp}$ is the component of the junction point displacement that is normal to the interface. We start with the A-block in the A-rich domain, and the B-block in the B-rich domain, such that both blocks behave approximately as ideal Gaussian chains with chain-lengths $N_A$ and $N_B$ respectively. When the chain has escaped into the B-rich domain, the A block takes on a partially collapsed conformation.

Due to the planar geometry, we utilize the 1-dimensional analogs of the collective variables from the MD portion of this work, CV1 ($x_\textrm{jp}$) and CV1$^\prime$ ($x_\textrm{cmA}$) to conduct different FFS simulations. We place the first interface slightly outside of the A-rich domain, $x_1 > x_I$. We place additional interfaces further and further out from the interface, with the final interface located at a sufficient distance for the chain to be fully detached from the interface.

The transition rate for the complete transition is computed by accumulating the transition probabilities between each interface. The equation for the transition rate is,
\begin{equation}\label{eq:ffs-rate}
k=\Phi_0\prod_i k_{i,i+t}
\end{equation}
where $\Phi_0$ is the flux of trajectories across the first interface, and $k_{i,i+1}$ is the transition probability from interface $i$ to $i+1$. We define a free energy for the transition according to
\begin{equation}\label{eq:ffs-fe}
    \beta\Delta F= -\ln\frac{k}{\Phi_0}
\end{equation}
which is motivated by the Arrhenius relationship, $k=A\exp(-\beta\Delta F)$ where $A$ is an unknown kinetic prefactor. It is important to note that the rate $k$ is a physical observable and should be insensitive to the definition of the basin. For example, if the first interface is placed at a larger $x$, then $\Phi_0$ will necessarily decrease, but there will also be some interfaces omitted which will cause $\prod_i k_{i,i+t}$ to increase. The overall effect is for $k$ to remain constant. Our definition of $\beta \Delta F$ will shift with the placement of the first interface, but not by enough to impact the scaling behavior.

\begin{figure*}[hbt!]
    \centering
    \includegraphics[width=1.0\linewidth]{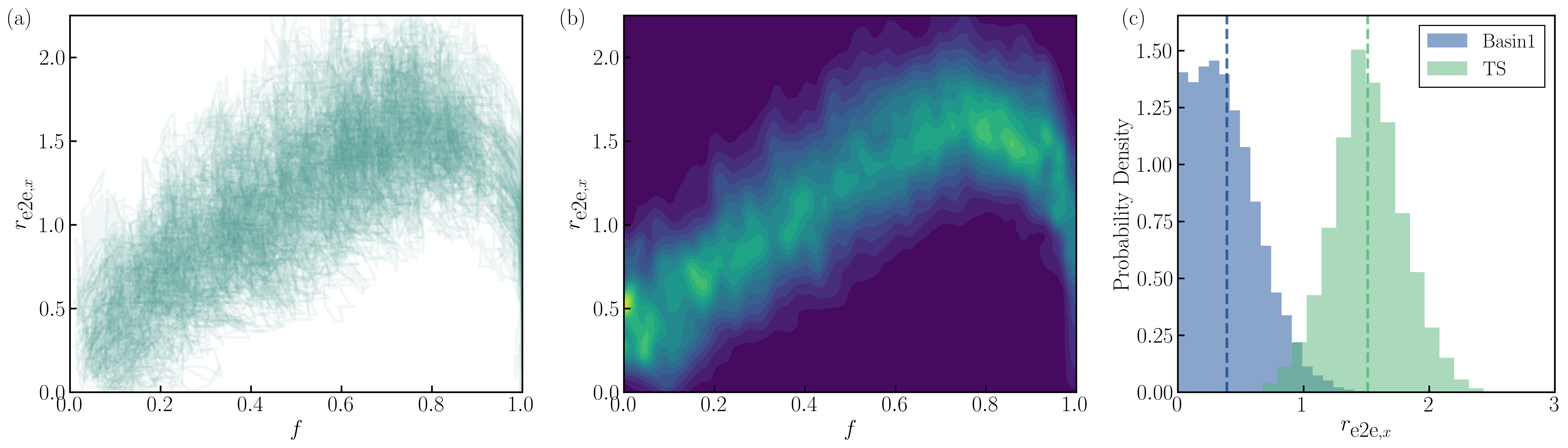}
    \vspace{-0.8cm}
    \caption{Results from the reactive ensemble for $N_A=N_B=16$, $z_c^{-1}=\epsilon=0.02$. (a) Example escape trajectories, (b) conditional probability distribution, $P(r_{\textrm{e2e},x}|f)$, where $r_{\textrm{e2e},x}$ is the $x$-component of the end-to-end distance of the core block and $f$ is the fraction CV defined in Equation \eqref{eq:frac}, and (c) the distributions of $r_{\textrm{e2e},x}$ in Basin1 and just before detaching from the interface, denoted as the transition state (TS). Vertical dashed lines mark the means of the two distributions.}
    \label{fig:e2e-hist}
\end{figure*}

\subsubsection*{Results and Discussion}
First, we compute the transition rate for the chain to escape from the interface using either the position of the junction point, $x_\textrm{jp}$, or the position of the center of mass of the A block, $x_\text{cmA}$, as the generating CV. We run FFS simulations for several core-block lengths, $N_A\in\{6,8,10,12,14,16,18\}$, to elucidate the scaling relationship of the free energy barrier. Figure \ref{fig:jun-rates} shows the free energy profiles and scaling behavior for the $x_\textrm{jp}$ CV. A complementary plot is provided for the $x_\text{cmA}$ CV in the ESI$^\dagger$.

The scaling of the free energy barrier with core-block length appears to be linear, regardless of the choice of generating CV. The inset of Figure \ref{fig:jun-rates}b also shows that the rate computed from Equation \eqref{eq:ffs-rate} decays exponentially with $N_A$. This is the expected scaling for the case when the core block does not collapse into a dry globule upon entering the B domain. Since a linear scaling of the barrier is also observed in experiments, this indicates that the experimental conditions could be such that the core block of an isolated chain is not fully collapsed after escaping. This would occur if the core block is too short to form a statistically probable globule. As argued by Lund et al., it could also occur if the corona block is sufficiently dense such that the core block cannot collapse fully upon entering the corona domain.\cite{lund_equilibrium_2011} As argued by Choi et al., it could also be due to solvent penetrating the collapsed globule such that all or most of the monomers are contacting the solvent.\cite{choi_molecular_2011,wu_globule--coil_1998,wang_comparison_1998} In our FFS simulations, the linear scaling is expected since the core block shrinks only slightly upon entering the B domain due to the high value of $\bar{\mathcal{N}}$.

Note that even after the chain has fully detached from the interface the free energy continues to gradually increase. This behavior is perfectly explained using the Markov chain for symmetric diffusion on a number line with an absorbing boundary condition on the left. After the chain has escaped, the chain still has a probability to diffuse backwards and fall into the starting basin. This probability decreases the further the chain is from the transition state. If we consider the absorbing boundary to be at the transition state, which we define as node 0, then we can compute the probability of reaching node $n+1$ starting from node $0$ before falling back into the starting basin; we define this probability as $P(n+1|0)=P_0$, and more generally we define $P(n+1|i)=P_i$ for $i\ge 0$. Starting from $P_0$ we can recursively compute all $P_i$ up to $P_n$. Lastly, we assume that the forward and backwards transition probabilities are both $1/2$ at all interfaces. By induction, we find that $P_n=P(n+1|n)=(n+1)/(n+2)$, and therefore $P_0 \sim 1/(n+2)$ and $-\ln P_n \rightarrow 0$ as $n\rightarrow\infty$. We could subtract out this contribution to obtain a flat plateau, however, we elect not to since the transition state is not well-defined for the monotonic free energy curves in Figures \ref{fig:jun-rates}. The small contribution does not affect the scaling of the free energy with $N_A$.

\subsubsection*{Reactive Ensemble (RE)}

From this point, it is of interest to evaluate the distributions of different chain properties along the trajectories to better understand the mechanism by which the chain is able to escape. We analyze different ensembles of trajectories specifically for the system with $N_A=N_B=16$, and $\epsilon=z_c^{-1}=0.02$. The forward flux ensemble (FFE) is the ensemble containing properties on the FFS interfaces for chains that were frozen immediately after reaching an interface. The FFE is simple to compute and analyze since polymer configurations on the interfaces are saved during the FFS simulation. However, it is hard to extract meaning from this ensemble due to the bias introduced through the \textit{first-crossing} condition. Therefore, we provide the FFE in the ESI$^\dagger$ using $x_\textrm{jp}$ and $x_\textrm{cmA}$ and simply note that these ensembles indicate chain extension during escape. The amount of extension present in the FFE is exaggerated compared to reality due to the uniqueness of the \textit{first-crossing} condition. In practice, it is more appropriate and meaningful to look at the chain properties from the ensemble of chain trajectories that successfully completed the transition between the starting and ending basin, with monomer coordinates written at regular intervals, as opposed to only at first-crossing. We denote this ensemble of transition paths as the \textit{reactive ensemble} (RE), which we discuss in this section.

The \textit{reactive} trajectories can be constructed by starting from chain conformations at the final interface and tracing them back to the first interface. These trajectories may cross each interface multiple times before reaching the final basin and may share common ancestors at intermediate interfaces. The only constraint on the reactive trajectories is that the chain must not fall back into the starting basin before reaching the ending basin.

Since the coordinates are output at regular intervals, the properties along the trajectories can be binned using any choice of collective variable (CV), which we denote as the \textit{selecting} CV. To track the progress of chain escape, we define a selecting CV based on the fraction of core-block monomers that have crossed the interface. To make this CV continuous, we employ a hyperbolic tangent switching function,
\begin{equation}\label{eq:frac}
f=\frac{1}{2N_A}\sum_{i=1}^{N_A}\left[1+\tanh\left(-\frac{x_i-x_I}{c}\right)\right],
\end{equation}
where $c=0.1$ modulates the width of the function, and $x_I$ is the interface position. By definition, $f\in[0,1]$. Example trajectories up to the point of detachment are shown in Figure \ref{fig:e2e-hist}a, which indicate that the primary mechanism for a chain to cross the interface is through extension.

The full ensemble of trajectories can generate any univariate or multivariate probability distribution. Here, we focus on $P(r_{\textrm{e2e},x},f)$ and, more specifically, the conditional distribution $P(r_{\textrm{e2e},x}|f)$, plotted in Figure \ref{fig:e2e-hist}b. This distribution quantifies the degree of chain stretching (normal to the interface) as a function of chain expulsion during successful escape attempts. Figures \ref{fig:e2e-hist}a and \ref{fig:e2e-hist}b show that the main escape pathway requires chains first to extend into solution (increasing both $f$ and $r_{\textrm{e2e},x}$) and then fully collapse and detach from the interface (increasing $f$ while decreasing $r_{\textrm{e2e},x}$). The conditional probability distribution seems to feature a 

To clarify this pathway and the stretched transition state, we compare the probability distributions $P(r_{\textrm{e2e},x}|f_{0})$ and $P(r_{\textrm{e2e},x}|f_{TS})$ in Figure \ref{fig:e2e-hist}c. The term "transition state" is not used formally here, but instead is simply used to indicate the point along the trajectory where $f$ becomes $1$. In other words, we define the "transition state" to be the step when the final bead crosses over the interface. The distributions in Figure \ref{fig:e2e-hist}c indicate the degree of extension for chains localized at the interface or actively detaching from it, respectively, confirming that the chains adopt an extended conformation immediately prior to detachment.

Our enhanced sampling MD simulations revealed that "bead-by-bead" extension of the chain into the solvent is a relatively low free energy pathway, even in the case where the core block collapsed into a dry globule. However, in that case the MFEP was still the HA mechanism. In the high density limit, the FFS simulations revealed that the chain escapes almost exclusively through an extension mechanism. This hints at the possibility that the MFEP crosses over from the HA mechanism (lower pathway on FES) to the stretching mechanism (upper pathway on FES) as the propensity for core collapse decreases.

\section{Conclusion}

In this study, we utilized coarse-grained molecular dynamics with spectral ABF enhanced sampling to compute the 2d free energy surface for the escape of a copolymer chain from a micelle at high segregation strength, and relatively low density. Our use of two collective variables ensured that the chain conformation was properly sampled during the escape process, and allowed us to observe a bimodal distribution in the chain conformation. In particular, near the transition state, we found that the chain can readily take either a collapsed or an extended conformation where some of the polymer beads remain in the micelle core. While the two conformations have commensurate free energies, they are separated by a large free energy barrier, further indicating the need for explicit biasing of the chain conformation. We computed the minimum free energy path using the string method on the free energy landscape and found it to be in agreement with the originally proposed Halperin--Alexander mechanism, both qualitatively and quantitatively. Namely, the MFEP featured a collapsed core block at the transition state that ultimately yielded a free energy barrier that scaled as $N_A^{2/3}$. Our 2d FES using the $R_\textrm{cm-jp}$ CV featured a broad region where the chain is still tethered to the micelle with a portion of the core block extended into the solution, indicating a high propensity for the extended conformation to exist. While these results are compelling in their validation of the Halperin--Alexander mechanism, they do not agree with experimentally observed scaling relationships of several TR-SANS experiments. This provides support for the idea that the core block is not fully collapsed during escape, which prompted us to study the escape mechanism for chains in a lamellae-forming diblock copolymer melt.

To compute the transition pathway in the high-density limit we conducted single-chain Monte Carlo simulations with forward flux sampling. We utilized a simplified model with a single bead-spring copolymer chain immersed in a mean-field background containing a sharp interface. We computed the rate and free energy barrier for escape using millions of independent forward flux sampling trajectories. The resulting free energy barrier scaled linearly with $N_A$ for different choices of collective variables. The linear scaling is in good agreement with experimental observations from TR-SANS measurements. We analyzed the properties of the polymer chains along the escape trajectories using the forward flux ensemble and the reactive ensemble and found that a large majority of the chains escape via an extended conformation, rather than the collapsed conformation of the Halperin--Alexander mechanism.

In this work we provided a detailed analysis of the different possible mechanisms for a diblock copolymer chain to escape from a micelle using two different simulation techniques. While the simulations were restricted to a narrow range of parameters, we were able to identify different pathways and their relative importance to chain exchange. In a future work we plan to use forward flux sampling simulations to do a more comprehensive study of the effect of the chain and matrix properties on the escape rate and mechanism.

\begin{acknowledgments}
We would like to thank Dr. Benjamin Ye for helping develop the custom potentials required to run our MD simulations within OpenMM. We would also like to thank Dr. Shensheng Chen for many helpful discussions. K.D.D. acknowledges the gracious hospitality of the California Institute of Technology during portions of this work. S.V. is supported by the U.S. Department of Energy, Office of Science, Office of Advanced Scientific Computing Research, Department of Energy Computational Science Graduate Fellowship under Award Number DE-SC0022158. Partial support for this research is provided by Hong Kong Quantum AI Lab, AIR@InnoHK of Hong Kong Government.
\end{acknowledgments}

\section*{Data Availability Statement}
The data that support the findings of this study are available from the corresponding author upon reasonable request.

\section*{References}
\bibliography{references,references-2}

\end{document}


\preprint{AIP/123-QED}

\def\*#1{\mathbf{#1}}
\def\&#1{#1}

\title{Supplementary Information for \textit{Mechanisms of Chain Exchange in Diblock Copolymer Micelles}}

\author{Samuel Varner}
\affiliation{Division of Chemistry and Chemical Engineering, California Institute of Technology, Crellin Laboratory, Pasadena, CA 91125, United States}

\author{Marcus M\"{u}ller}
\affiliation{Institut für Theoretische Physik, Georg-August-Universität Göttingen, Friedrich-Hund-Platz 1, D 37077 Göttingen, Germany}

\author{Alejandro Gallegos}
\affiliation{Department of Chemical and Materials Engineering, New Mexico State University, Las Cruces, New Mexico 88001, United States}

\author{Timothy P. Lodge}
\affiliation{Department of Chemical Engineering \& Materials Science, University of Minnesota, Minneapolis, Minnesota 55455, United States}
\affiliation{Department of Chemistry, University of Minnesota, Minneapolis, Minnesota 55455, United States}

\author{Kevin D. Dorfman}
\affiliation{Department of Chemical Engineering \& Materials Science, University of Minnesota, Minneapolis, Minnesota 55455, United States}

\author{Zhen-Gang Wang}
\affiliation{Division of Chemistry and Chemical Engineering, California Institute of Technology, Crellin Laboratory, Pasadena, CA 91125, United States}
\email{zgw@caltech.edu}

\date{\today}
\maketitle

\subsubsection*{MD Properties}

\begin{figure}[h!]
    \centering
    \includegraphics[width=0.6\linewidth]{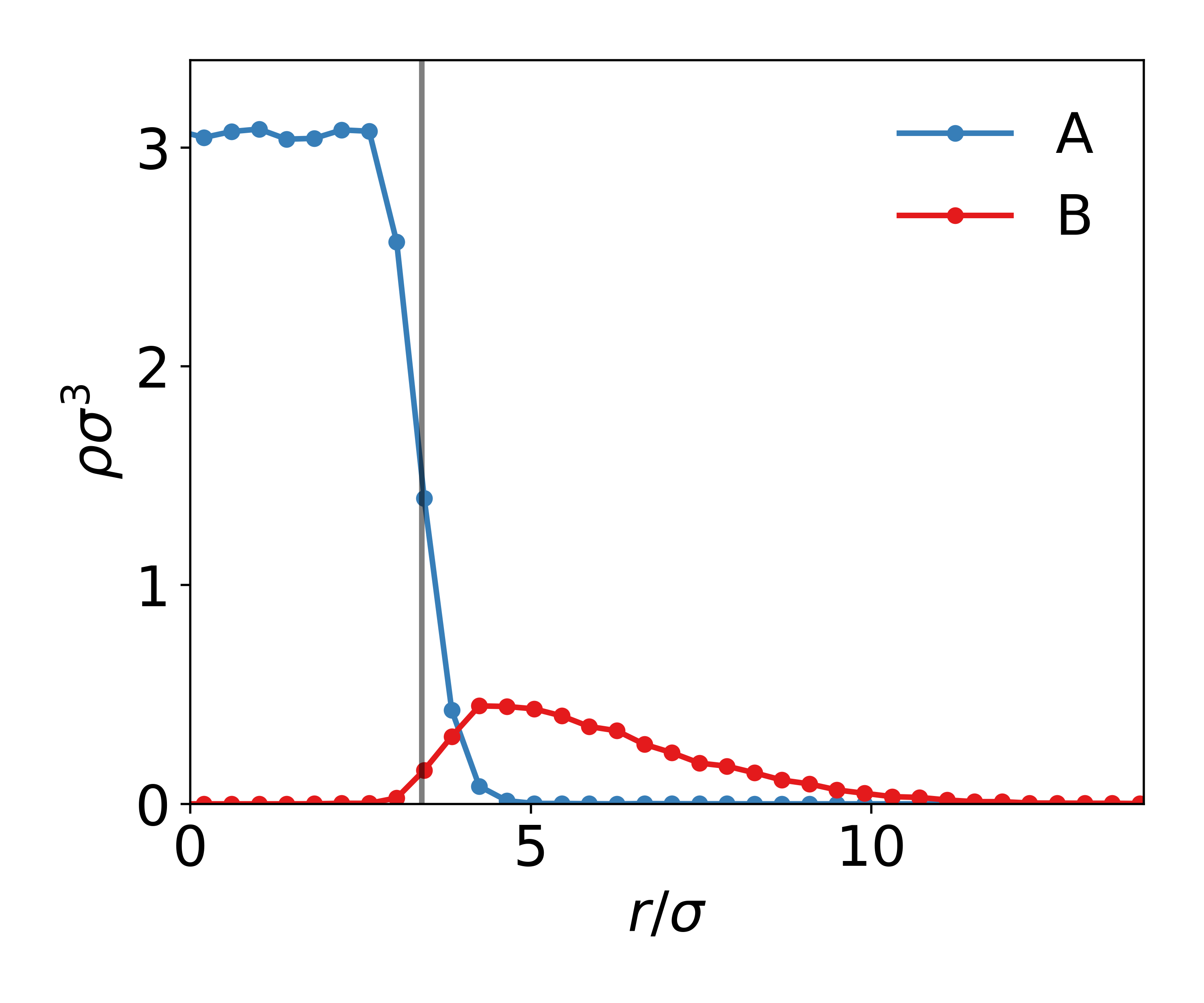}
    \vspace{-1.1cm}
    \caption{Radial density profile of the micelle computed from MD for $N_A=15$, $N_B=24$, $\beta\Delta\epsilon=23$. The vertical gray line corresponds to the average value of the junction point location $\langle R_{cm-jp} \rangle$ as computed from FES1 in Figure 2.}
    \label{fig:1dprofile}
\end{figure}
\begin{figure}[h!]
    \centering
    \includegraphics[width=0.6\linewidth]{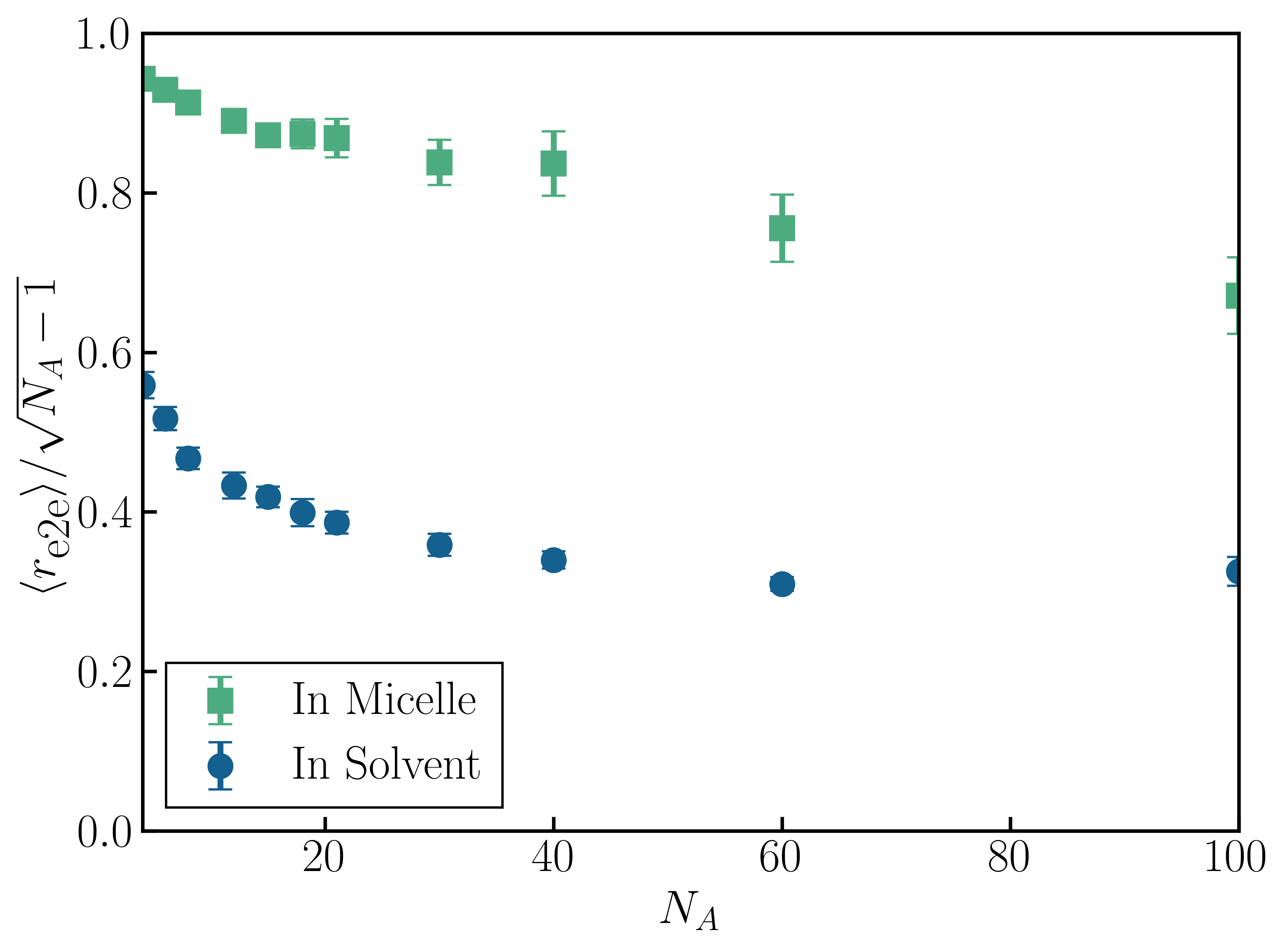}
    \vspace{-0.8cm}
    \caption{Average end-to-end distance of the core block when inside the micelle versus in the solvent. The end-to-end distance deviates from ideal when inside the micelle for large $N_A$ because the core constrains the chain extension slightly.}
    \label{fig:polymersizes}
\end{figure}

\subsubsection*{String Method}

With our 2d FES in hand, we utilize the string method to compute the MFEP of the expulsion process.\cite{e_string_2002,e_simplified_2007,e_transition-path_2010} In the string method, images are placed between Basin1 and Basin2 in a systematic way (usually linearly interpolated). The path connecting the images defines the string. The string is relaxed by updating the position of the images using the gradient of the FES as the driving force. Importantly, the string of images has to be reinterpolated after each iteration to prevent the images from all falling into one of the two basins. Once the string has converged and become stationary, the string represents the MFEP. Given a 2d FES $F(x,y)$ and a starting string of images $\phi_0(\*x,\*y)$, the MFEP is found through an iterative process. First the positions of the images are updated,
\begin{equation}
    \hat\phi_1(\*x,\*y) = \phi_0(\*x,\*y) - \lambda\nabla F(\*x,\*y)
\end{equation}
followed by a reinterpolation of the images along the string at equal arc-length intervals, $\hat\phi_1(\*x,\*y)\rightarrow\phi_1(\*x,\*y)$. The process is finished when $||\phi_{i}-\phi_{i-1}||_2 < \varepsilon$ where $\varepsilon$ is some tolerance.

\subsubsection*{Free Energy Projections}

In our MD simulations we choose to bias 2 collective variables simultaneously in order to more efficiently sample the conformation of the chain at each point along the pseudo reaction coordinate. It is of interest to compute the 1d free energy profiles in the pseudo reaction coordinates $R_\textrm{cm-cm}$ and $R_\textrm{cm-jp}$. In this section, we denote the reaction coordinate variable as $R$ and the conformation variable as $r$. To project the free energy surface onto $R$, we first compute the probability distribution.
\begin{equation}
    P(R,r)=\exp[-\beta F(R,r)]
\end{equation}
We then compute the 1d probability distribution by integrating over $r$.
\begin{equation}
    P(R)=\int dr P(R,r)
\end{equation}
The 1d free energy profile is then simply computed from the 1d probability distribution.
\begin{equation}
    \beta F(R)=-\ln P(R)
\end{equation}
In principle, a simulation where only $R$ is biased would produce the same $\beta F(R)$ with perfect sampling. However, barriers in $r$ would prevent perfect sampling of the conformation. As a result, we expect the 1d profiles computed from the projection of the 2d free energy surface to be more reliable.

\subsubsection*{Forward-flux Sampling}



For the degree of segregation used in this study it is highly unlikely for a polymer to escape from a micelle. Thus, we employ forward-flux sampling (FFS)\cite{allen_forward_2006,borrero_reaction_2007,hussain_studying_2020} to enhance the transition rate of the polymers, and to get an accurate estimate of the rate constant specifically in the case of our single-chain MC simulations. FFS is a transition path sampling (TPS) technique that can be used to compute the rate of a rare barrier-crossing event and even obtain reactive trajectories corresponding to the ensemble of possible kinetic pathways. FFS consists of placing virtual interfaces between a starting basin (Basin1) and an ending basin (Basin2) along some reaction coordinate. The interfaces are used as \textit{checkpoints} to save configurations of trajectories which have reached a given interface, such that the trajectories can be restarted from those points when they inevitably fall into one of the two basins. If we have a set of $N_i$ configurations that were saved at interface $i$, then the FFS algorithm amounts to selecting $M$ configurations out of $N_i$ and continuing each trajectory until they either reach the next interface, or fall back into the basin. This is continued until a minimum threshold of configurations $N_{i+1}$ reach the next interface. If $M$ is the number of required trajectories to reach $N_{i+1}$ \textit{successes}, then the transition rate between interface $i$ and $i+1$ can be computed as:
\begin{equation}
    k_{i,i+1} = \frac{N_{i+1}}{M}
\end{equation}
with the free energy change being,
\begin{equation}
    \beta\Delta F_{i,i+1} = -\ln k_{i,i+1}
\end{equation}
Additionally, we need to estimate the flux from the starting Basin1 across the first interface, $\Phi_0$, which can be computed by running a long simulation in Basin1, and monitoring the rate at which the particle crosses over the interface in the forward direction. The rate is computed as, 
\begin{equation}
    \Phi_0=\frac{N_0}{\tau}
\end{equation}
where $N_0$ is the number of forward crossings, and $\tau$ is the total simulation time. Each of the $N_0$ crossing configurations can be saved and used in the next step of the FFS algorithm to propagate from interface 0 to interface 1. The full rate constant for the transition from Basin1 to Basin2 is computed as,
\begin{equation}
    k=\Phi_0\prod_i k_{i,i+1}
\end{equation}
and we define the total free energy change is $\beta \Delta F = -\ln (k/\Phi_0)$. As mentioned in the main text, this is simply a definition, and it's value will depend on the definition of the starting basin, whereas the overall rate $k$ will be insensitive to such changes.

\subsubsection*{Monte Carlo Simulation}

We run millions of embarrassingly parallel single-chain Monte Carlo simulations (MCS) using the Hamiltonian described in the main paper, which we denote as $H$. The MC steps are taken in the following manner:
\begin{enumerate}
    \item The initial value of the Hamiltonian is computed, $H_v$
    \item A single bead is moved by drawing from an independent random Normal distribution for each Cartesian coordinate. The Normal distributions have mean 0 and standard deviation 1.
    \begin{align}
        x_{v'} &= x_{v} + (\sigma/\sqrt{N-1})\pmb{\hat{\mathcal{N}}}(0,1) \\[5pt]
        y_{v'} &= y_{v} + (\sigma/\sqrt{N-1})\pmb{\hat{\mathcal{N}}}(0,1) \\[5pt]
        z_{v'} &= z_{v} + (\sigma/\sqrt{N-1})\pmb{\hat{\mathcal{N}}}(0,1)
    \end{align}
    where $\pmb{\hat{\mathcal{N}}}(0,1)$ is the standard normal distribution from the PCG family of generators.\cite{oneill:pcg2014} We set $\sigma=1$ for simplicity. Note that these moves are used to reproduce Rouse dynamics.
    \item The new value of the Hamiltonian is computed, $H_{v'}$.
    \item The Metropolis-Hastings algorithm is used to accept or reject the move. Namely, the move is accepted if $\beta(H_{v'}-H_v) = \beta\Delta H_{v,v'} \le 0$. Additionally, if $\beta(H_{v'}-H_v) = \beta\Delta H_{v,v'} > 0$, then the move is accepted with probability $\exp(-\beta \Delta H_{v,v'})$. If neither condition is satisfied, then the move is rejected, and microstate $v$ is restored.
\end{enumerate}
Each polymer is treated independently and therefore has a separate Hamiltonian.

\subsubsection*{Forward Flux Ensembles}
Here we show the forward flux ensemble (FFE) computed for $N_A=N_B=16$ and $\epsilon=z_c^{-1}=0.02$ for both generating cvs ($x_\textrm{jp}$ and $x_\textrm{cmA}$)
\begin{figure}[h!]
    \centering
    \includegraphics[width=1\linewidth]{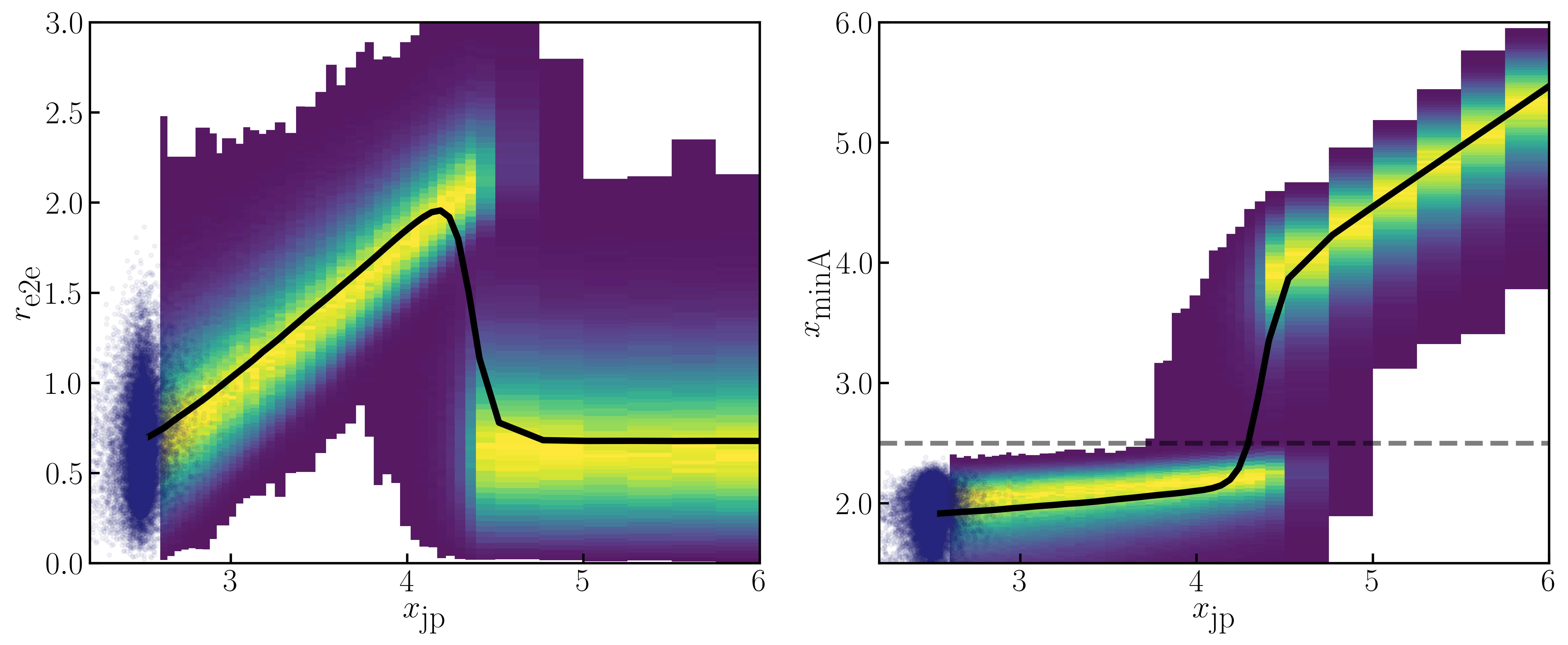}
    \vspace{-1.3cm}
    \caption{FFE of the (left) core-block end-to-end distance and (right) core-block minimum bead position, as a function of the chain junction point. Here, $x_\text{jp}$ is used as the generating CV.}
    \label{fig:jun-e2e}
\end{figure}
\begin{figure}[h!]
    \centering
    \includegraphics[width=1\linewidth]{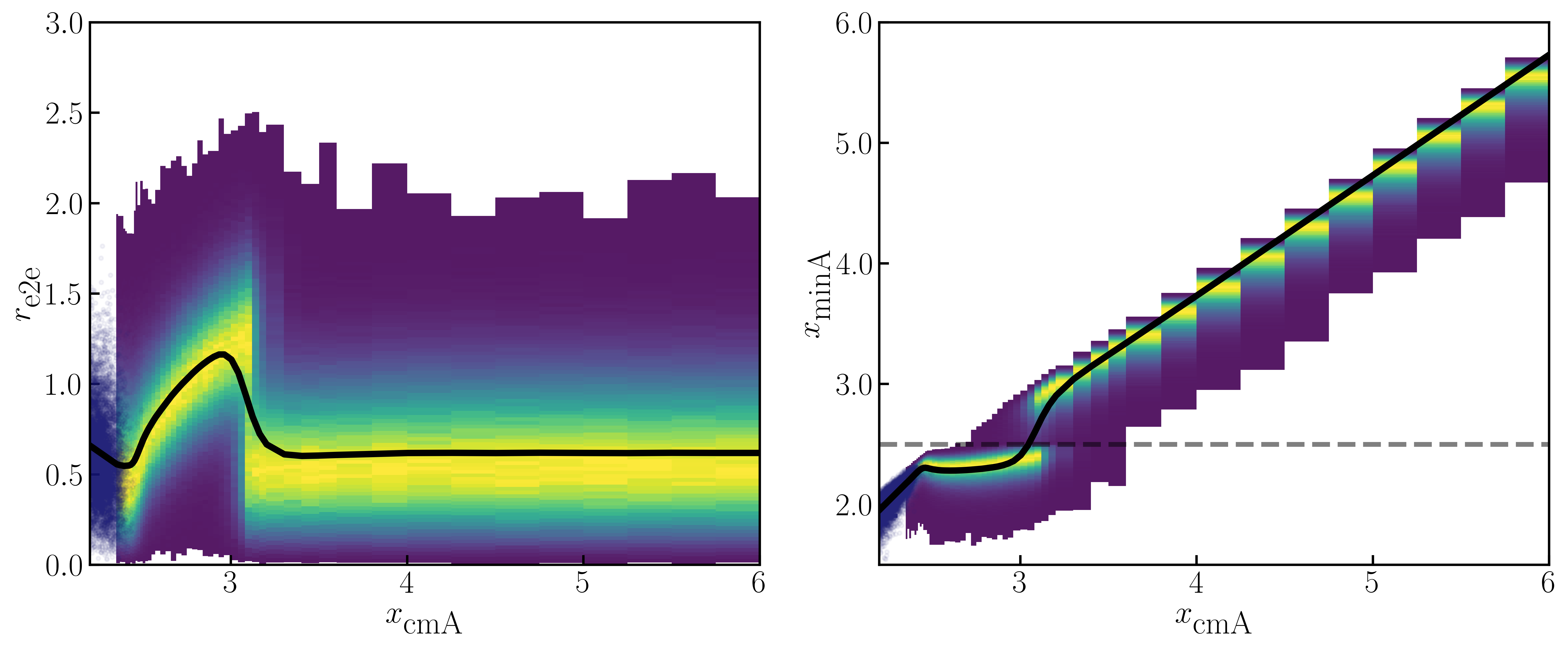}
    \vspace{-1.3cm}
    \caption{FFE of the (left) core-block end-to-end distance and (right) core-block minimum bead position, as a function of the core-block center-of-mass. Here, $x_\text{cmA}$ is used as the generating CV.}
    \label{fig:comA-e2e}
\end{figure}

\subsubsection*{Free Energy and Rate using Center of Mass CV}
\begin{figure*}[h!]
    \centering
    \includegraphics[width=1\linewidth]{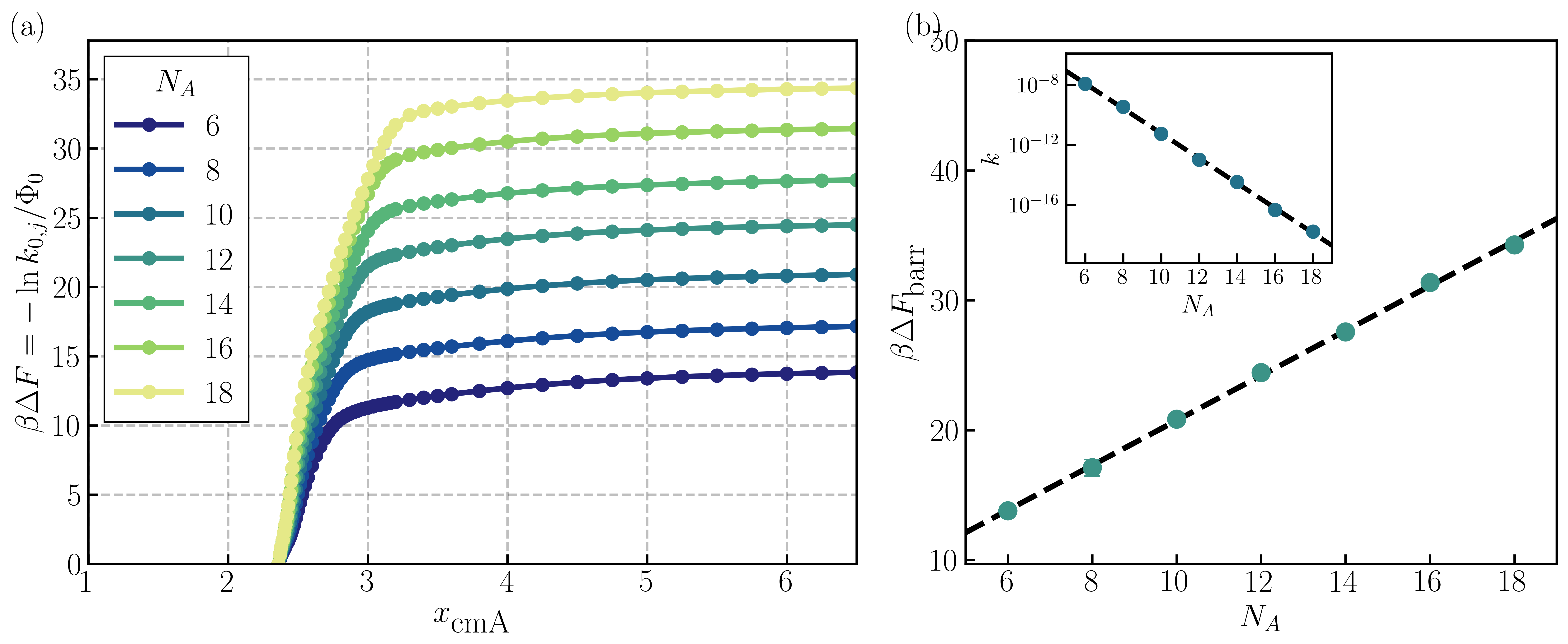}
    \vspace{-1.5cm}
    \caption{FFS results for the $x_\text{cmA}$ generating CV with various core-block lengths. (left) The natural logarithm of the cumulative transition rate. (right) The total free energy change versus core-block length, with optimized linear fit. Error bars represent a 95\% confidence interval from averaging 4 simulations.}
    \label{fig:comB-rates}
\end{figure*}

To further validate the FFS simulations and ensure that our results were general and not specific to any choice of collective variable, we recomputed some of the main results using the center of mass of the core (A) block, denoted as $x_{cmA}$. Figure \ref{fig:comB-rates} shows that the scaling of the free energy barrier is linear in the core block length, even when using the alternative CV.

\newpage

\subsubsection*{Property Distributions from Forward Flux Ensemble}

\begin{figure}[t!]
    \centering
    \includegraphics[width=0.95\linewidth]{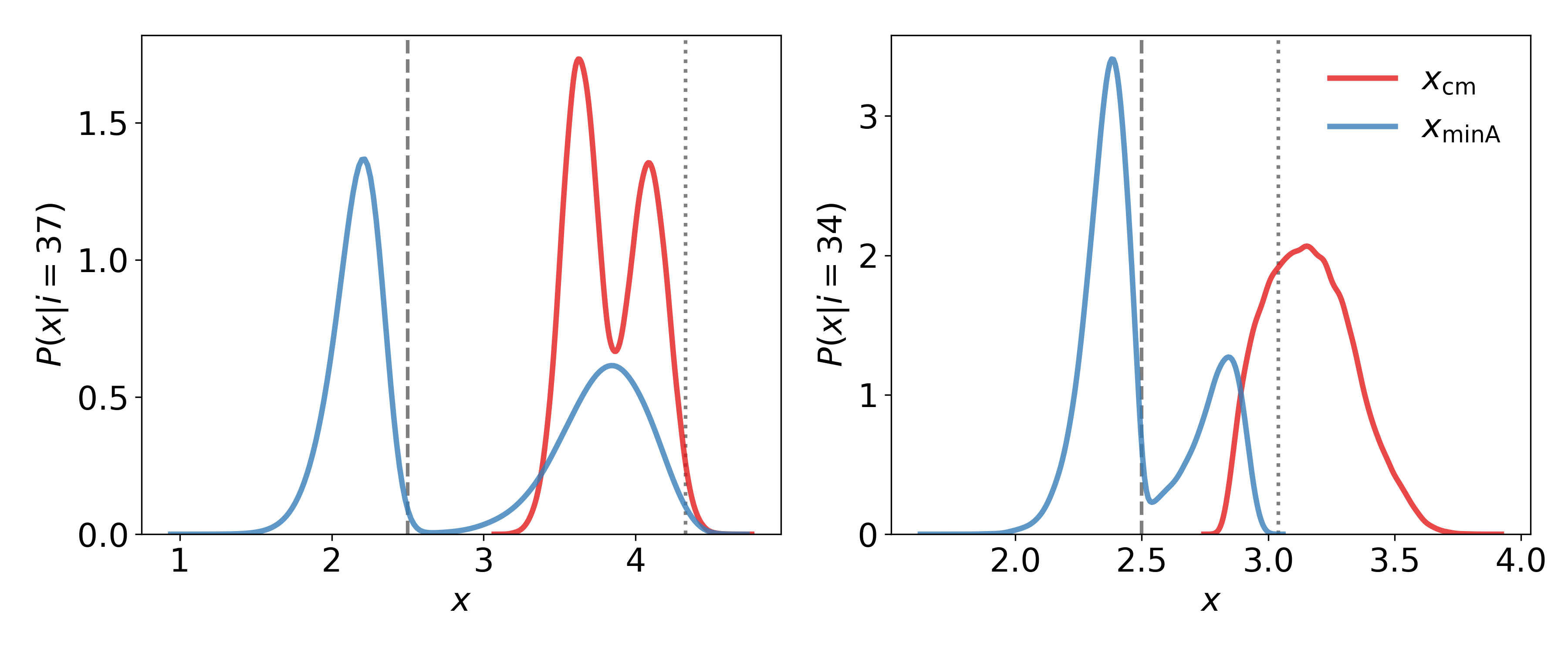}
    \vspace{-1.1cm}
    \caption{Distribution of chain properties at the transition state from FFE for (left) $x_{jp}$ and (right) $x_\text{cmA}$ generating CVs. The dashed line is the location of the interface. The dotted line is the value of the corresponding generating CV at the selected interface.}
    \label{fig:ffe-dist}
\end{figure}

\begin{figure}[h!]
    \centering
    \includegraphics[width=0.95\linewidth]{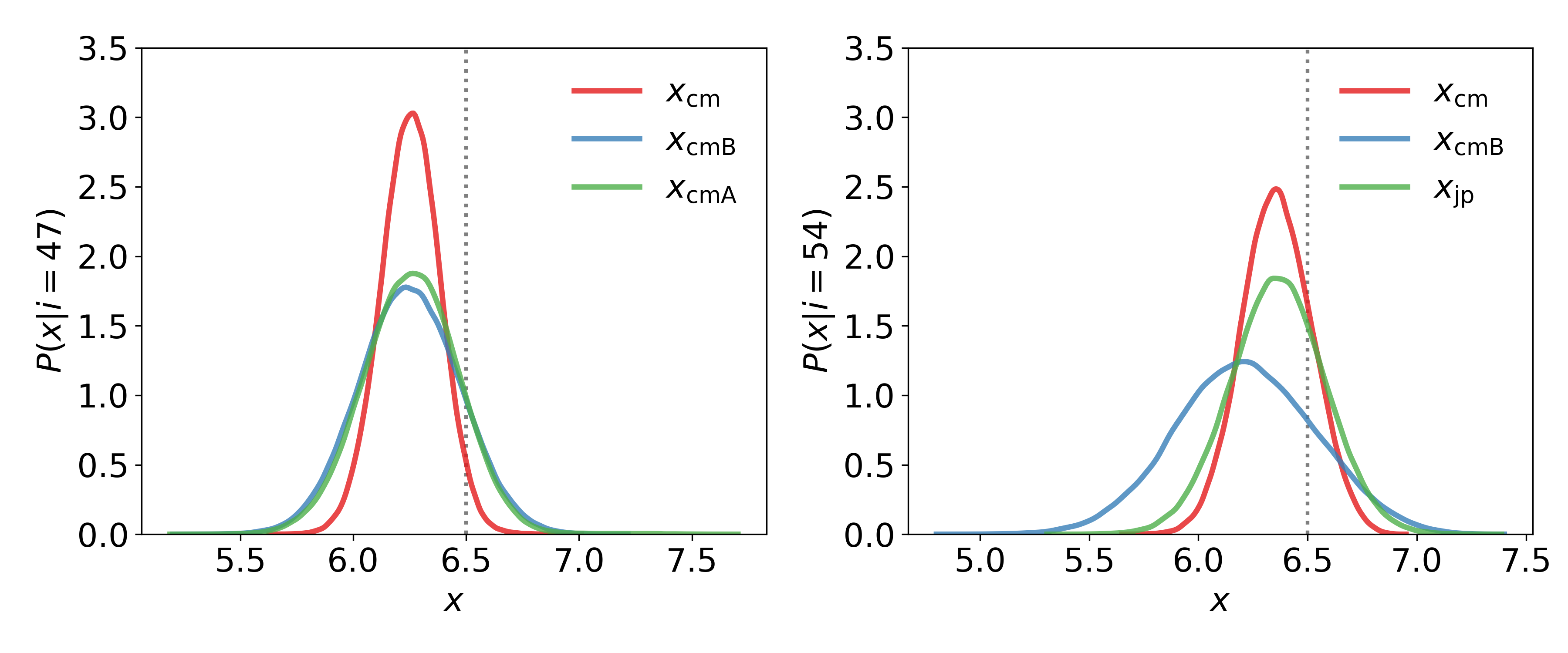}
    \vspace{-1.1cm}
    \caption{Distribution of chain properties at the final interface for (left) $x_{jp}$ and (right) $x_\text{cmA}$ generating CVs. The dotted line is the value of the corresponding generating CV at the selected interface.}
    \label{fig:ffe-dist-final}
\end{figure}

We also plot the distribution of chain properties at the transition state for the $x_{jp}$ and $x_\text{cmA}$ generating CVs in Figure \ref{fig:ffe-dist}. Here, we simply select an interface that contains a bimodal distribution in $x_\text{minA}$. The bimodal distribution indicates that for that particular value of the generating CV, there are some chains which are extended and some which are escaped and collapsed. The left panel of Figure \ref{fig:ffe-dist} shows a bimodal distribution on $x_\text{minA}$ as well as $x_\text{cm}$ where the two modes correspond to the collapsed and extended conformations. The two conformations are also present in the right panel, however, the degree of extension is diminished, and the resulting $x_\text{cm}$ distribution is no longer bimodal. The different shapes and features of the distributions from different generating CVs are largely due to the nature of the first-crossing condition of the FFE. We also provide distributions for the final interface in Figure \ref{fig:ffe-dist-final} to verify that Basin2 is placed far enough from the transition state so as not to affect the FFS results. We require the distributions on the final interface to mimic those of a chain in an isotropic fluid with some inherent biasing due to the \textit{first crossing} condition of the FFS algorithm.


\newpage
\bibliography{references,references-2,pcg}